\documentclass{aa}  

\usepackage{graphicx}
\usepackage{natbib}
\usepackage{mathtools}
\usepackage{multirow}
\usepackage{amsmath}
\usepackage{url}
\usepackage{txfonts}

\graphicspath{{./}{figures/}}

\begin{document}

   \title{Observations of Fan-Spine Topology by Solar Orbiter/EUI: Rotational Motions and Indications of Alfvén Waves}


   \author{E. Petrova \inst{1}
        \and T. Van Doorsselaere\inst{1}
        \and D. Berghmans\inst{2}
        \and S. Parenti\inst{3}
        \and G. Valori\inst{4}
        \and J. Plowman\inst{5}
          }

   \institute{Centre for mathematical Plasma Astrophysics, Mathematics Department, KU Leuven, Celestijnenlaan 200B bus 2400, B-3001 Leuven, Belgium\\
        \email{elena.petrova@kuleuven.be}
         \and Royal Observatory of Belgium, Ringlaan -3- Av. Circulaire, 1180 Brussels, Belgium
        \and Université Paris-Saclay, CNRS, Institut d’Astrophysique Spatiale, 91405, Orsay, France
        \and Max-Planck-Institut für Sonnensystemforschung, Göttingen, Germany
        \and Southwest Research Institute, Boulder, CO 80302, USA\\
             }

   \date{Received ; accepted }

 
  \abstract
   {Torsional Alfvén waves do not produce any intensity variation and are, therefore, challenging to observe with imaging instruments. Previously, Alfvén wave observations were reported throughout all the layers of the solar atmosphere using spectral imaging. }
   {We present an observation of a torsional Alfvén wave detected in an inverted y-shape structure
   observed with the HRIEUV telescope of the EUI instrument onboard Solar Orbiter in its 174 $\AA$ channel. The feature consists of two footpoints connected through short loops and a spine with a length of 30 Mm originating from one of the footpoints.}
   { In the current work, we also make use of the simultaneous observations from two other instruments onboard Solar Orbiter. The first one is PHI (Polarimetric and Helioseismic Imager) that is used to derive the magnetic configuration of the observed feature. The second one is SPICE (Spectral Imaging of the Coronal Environment) that provided observations of intensity maps in different lines including Ne VIII and C III lines.
     We also address the issues of the SPICE point spread function and its influence on the Doppler maps via performed forward modeling analysis.}
    { The difference movie shows clear signatures of propagating rotational motions in the spine. We measure propagation speeds of 136 \mbox{km/s} - 160 \mbox{km/s} which are consistent with the expected Alfvén speeds. The evidence of rotational motions in the transverse direction with velocities of 26 \mbox{km/s} - 60 km/s serves as an additional indication of torsional waves being present. Doppler maps obtained with SPICE show strong signal in the spine region. Under the assumption that the recovered point spread function is mostly correct, synthesized raster images confirm that this signal is predominantly physical. }
    {We conclude that the presented observations are compatible with an interpretation of either propagating torsional Alfvén waves in a low coronal structure or untwisting of a flux rope. This is the first time we see signatures of propagating torsional motion in corona as observed by the three instruments onboard Solar Orbiter. }

\titlerunning{Observations of Fan-Spine Topology:  Rotational Motions and Indications of Alfvén Waves  }
\maketitle
%
\section{Introduction}
The capabilities of solar imaging instruments drastically advanced throughout the last couple of decades allowing to access unprecedented spatial and temporal resolution. Therefore, modern imaging instruments provide means for many theoretically predicted phenomena to be finally observed. 
The theory of MHD waves in a cylindrical geometry was developed several decades ago \citep{zaitsev1975}, and it predicted that there should be several wave modes with different characteristics in structures, like for example, coronal loops - magnetic tubes filled with plasma and rooted in the solar surface \citep{Reale2014}, that can be described by this formalism. 
Although observations of acoustic or kink waves are often reported \citep{Anfinogentov2015,Gao2022,Shrivastav2023}, there is still a lack of Alfvén wave observations. Ironically, it is one the potentially most interesting phenomena to observe due to its ability to carry energy from the lower layers of the atmosphere to the corona \citep{Soler2019}.  
In a flux tube, propagating Alfvén waves manifest themselves as rotational perturbations of the plasma velocity and azimuthal components of the magnetic field \citep{edwin1983}. Since the Alfvén waves are incompressible, they do not produce any intensity perturbations in the linear regime. They also do not exhibit displacement of the structure as, for example, kink waves, therefore making it quite challenging to observe with imaging instruments \citep{vd2008}. Hence, they are usually observed via spectral data using Doppler maps with the signature being red and blue shifts occurring simultaneously and periodic nonthermal line broadening \citep{Erdelyi2007}. However, when it comes to the interpretation of rotational motion observations, it is important to note that not only torsional Alfvén waves manifest themselves as rotations. It has been reported by \cite{Goossens2014} that the Doppler velocity field of kink waves, depending on the line of sight, can look quiet similar to what is expected from Alfvén waves. 

Due to their properties, once the Alfvén waves are generated, they can easily propagate along the magnetic flux tubes \citep{Morton2023}. There are numerous reports of Alfvén wave detection in different layers of the solar atmosphere, most of them used spectral observations such as, for example, line broadening. 
\cite{Stangalini2021} reported observations of torsional oscillations in the photosphere within a magnetic pore. A study of the evolution of the magnetic field with the use of the Interferometric Bidimensional Spectropolarimeter (IBIS) revealed the presence of two out-of-phase torsional oscillations in different lobes of the pore corresponding to an azimuthal waves number $m=1$. Inferred energies exhibited an energy flux of approximately 140 $\mathrm{kWm^{-2}}$ for both lobes. Corresponding numerical simulations suggest that a kink mode could be a possible excitation mechanism of these waves.

Rotating motions are also observed in chromospheric swirls, often referred to as "magnetic tornadoes \citep{Shetye2019}. Plasma from the chromosphere propagates upwards in a spiraling motion, therefore resulting in swirl signatures. This phenomenon have garnered significant interest due to their capability to transfer mass and energy into the corona. \cite{Wedemeyer2012} have shown that these swirls in the form of torsional Alfvén waves can provide a flux of 440 $\mathrm{Wm^{-2}}$ to the lower corona. It was numerically shown as well that indeed Alfén waves can be generated by vortex flow in the chromosphere \citep{fedun2011}. This energy supply is substantial enough to heat the quiet corona \citep{Withbroe1977} and thus presents a potential alternative for energizing the corona.
Small-scale swirls (with diameter up to 3 Mm) exhibit Doppler velocities of up to 7 $\mathrm{km/s}$ and are typically associated with the motions of magnetic concentrations in photosphere \citep{Wedemeyer2009}.

Observations of torsional waves in the chromosphere were also reported by \cite{Jess2009}. Using observations from the Swedish Solar Telescope (SST) that observed bright points in H$\alpha$ they detected variation of line width at full-width half-maximum (FWHM)  with periods from 126 to 700 seconds. Combining the energy that can be carried by similar bright points over the solar surface it would produce on average 240 $\mathrm{Wm^{-2}}$.
\cite{DePontieu2012} also used the data from SST to detect that spicules of type II exhibit three types of motions  - flows in a direction of the magnetic field, swaying motions, and torsional motions of the order of 25-30 km/s. Their observed torsional motions are present in most of the spicules and represent, according to the author's interpretation, Alfvénic waves with a speed of propagation of several hundred \mbox{km/s}. Similar observations for spicular-type structures exhibiting high-frequency ($<50s$ period) oscillations were made by \cite{Srivastava2017}.

\cite{Kohutova2020} reported the first direct observations of propagating torsional Alfvén waves at coronal height using observations from AIA SDO \citep{Lemen2012} and IRIS \citep{DePontieu2014}. The torsional oscillation was detected in the Doppler velocity signal as the opposite edges of the flux tubes showed an anti-phase oscillation. Over time the observed oscillations demonstrated an attenuation which can be explained by phase mixing. The propagation speed was estimated to be 170 km/s. It was also shown that the observed Alfvén waves were generated as a result of a magnetic reconnection. The authors additionally suggest that small-scale reconnection events happening in twisted magnetic elements could also excite Alfvén waves \citep{cirtain2007}. 

One of the common structures with reconnection happening has a particular form of so-called inverse Y shape \citep{Pariat2009}. It is often related to jets \citep{Yokoyama1995}, which footpoints are not a mere singular bright point, instead, they display a configuration reminiscent of a cusp or an inverted Y-shape. It was introduced as a result of the observations by \citet{Shibata1994, Shibata2007}. This shape is formed as a result of reconnection between a bipole and the ambient vertical field. Reconnection produces a massive jet and generates a flux of torsional Alfvén waves. 

\begin{figure}
  \resizebox{\hsize}{!}{\includegraphics{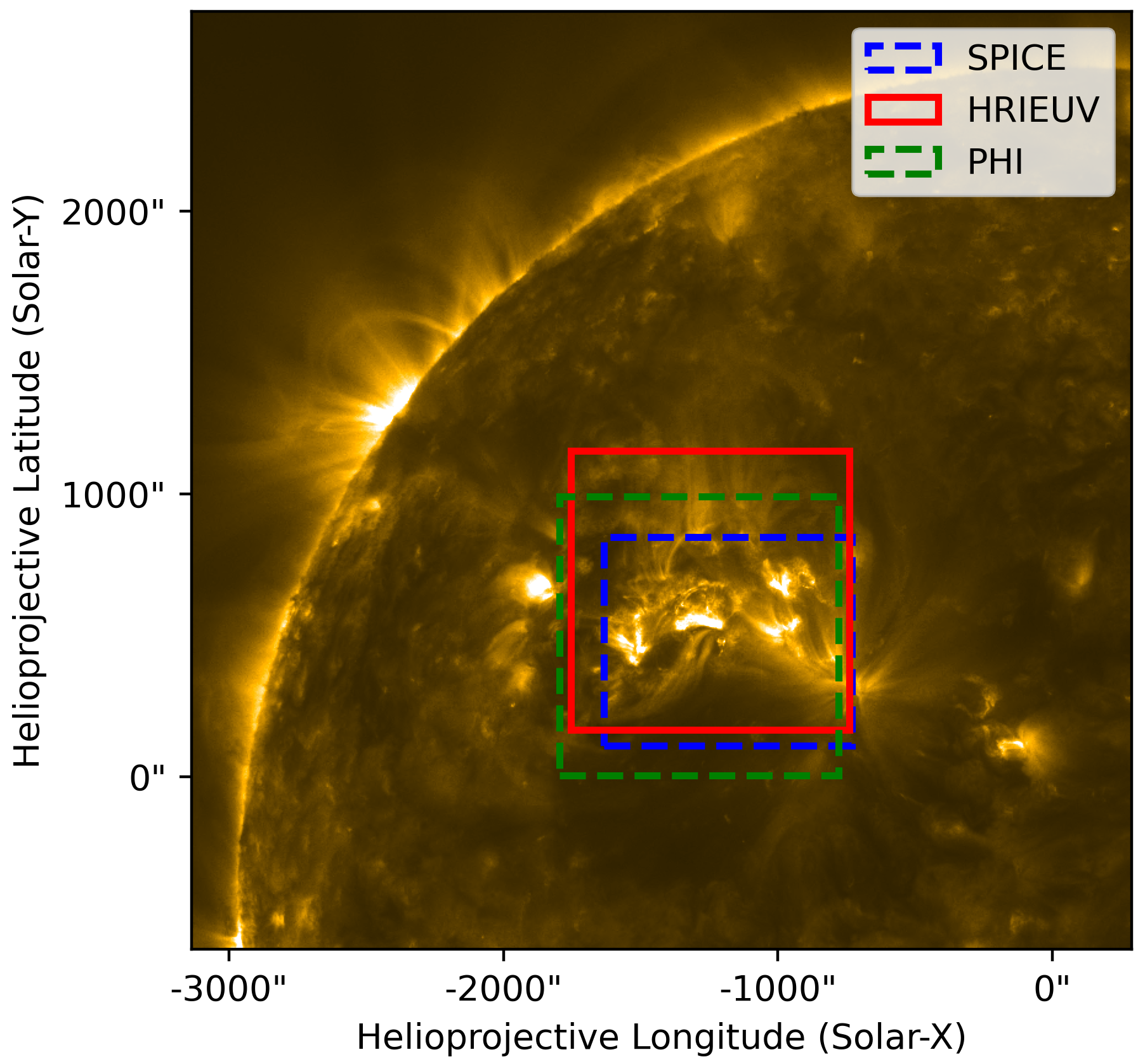}}
\caption{Overview of the instrument's field of views (FoV) as viewed from the Solar Orbiter FSI (Full Sun Imager) in 174 \AA. Overlaid on top are the HRIEUV FoV shown by red rectangle, SPICE and PHI FoVs are shown by blue and green rectangles respectively. 
\label{fig:fov}}
\end{figure}

\begin{figure}
 \resizebox{\hsize}{!}{\includegraphics{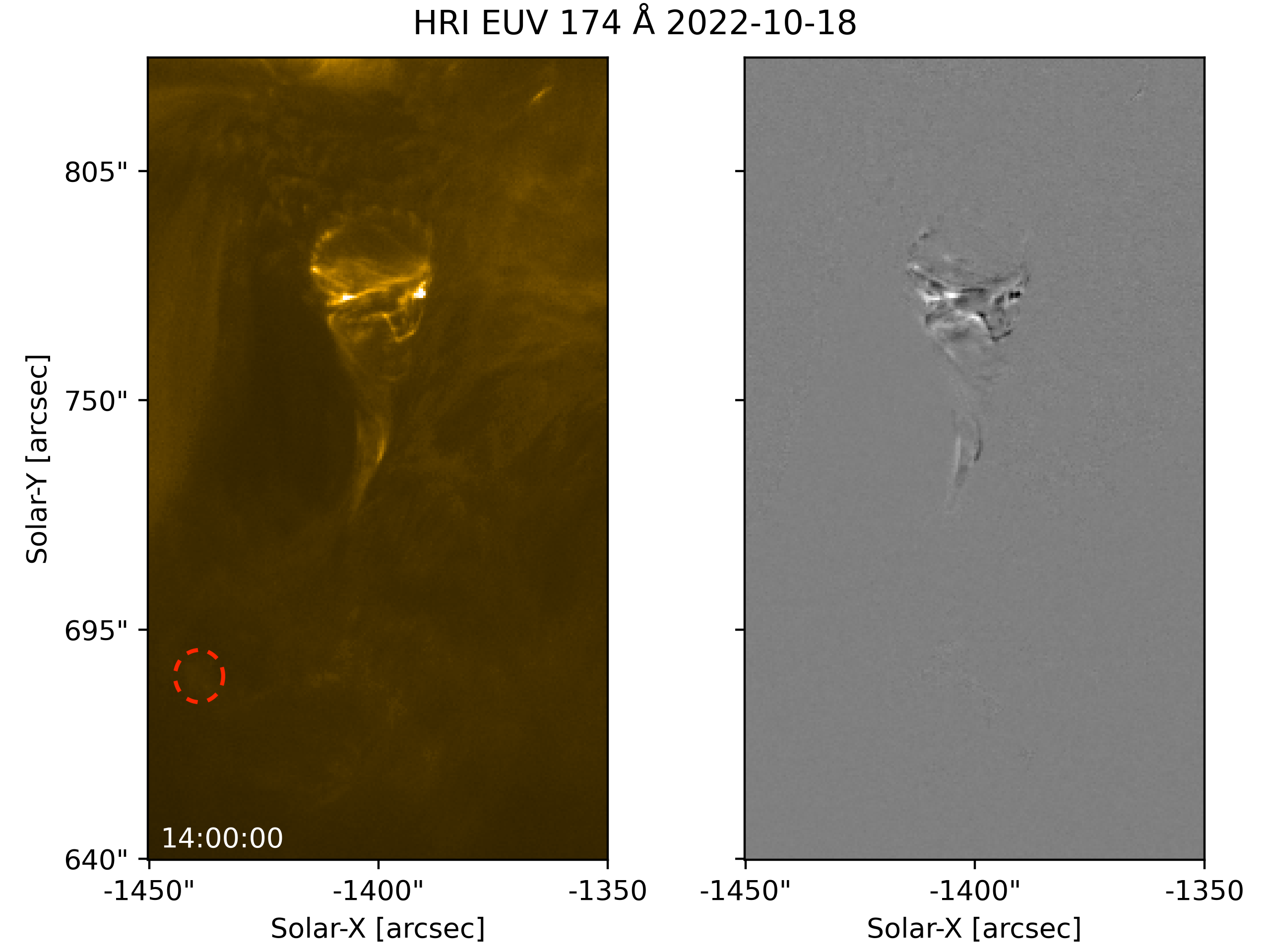}}
     \caption{The first frame of the sequence of images taken by HRIEUV. The left panel shows the original movie. The right panel shows a running difference movie mimicking panel on the left. Spine dynamics shows untwisting motions propagating towards the remote footpoint denoted by the red circle in the left panel. 
     \label{fig:video1}}
\end{figure}

\begin{figure*}[h!]
 \centering
   \includegraphics[width=17cm]{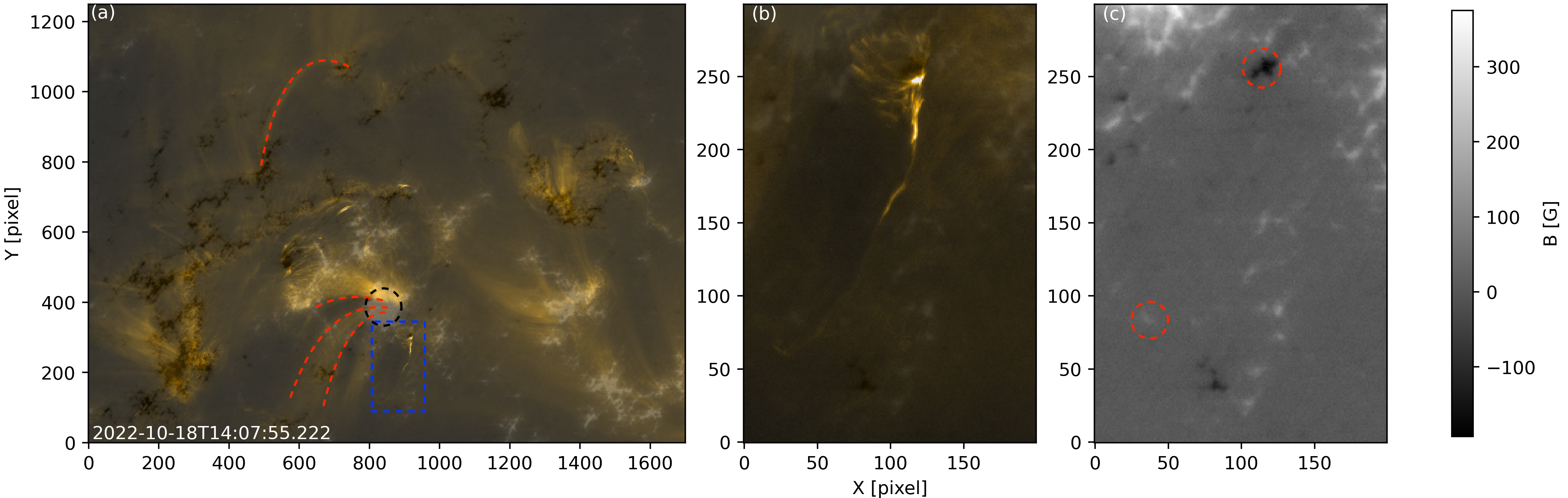}
     \caption{Magnetograms obtained with PHI HRT. Panel (a) shows the cropped field of view of HRIEUV imposed on top of the HRT view. The blue rectangle shows the location of the spine. Red dashed curves show identified loops used for alignment purposes. The black dashed circle shows a sunspot nearby the spine. Panel (b) shows a zoomed-in view of the feature. Panel (c) shows a PHI zoomed-in view alone. Red dashed circles show the footpoint at the origin of the spine and the footpoint corresponding to the remote brightening}
     \label{fig:PHI_full}
\end{figure*}

In a similar magnetic configuration, there is a so-called fan-spine topology which is formed in a similar fashion through the emergence of a magnetic bipole into a pre-existing unipolar field \citep{Yang2020}. It has been proven that this kind of topology is favorable for the occurrence of solar flares through null-point reconnection \citep{Cheng2023}. Some theoretical and numerical models have been proposed where in this configuration reconnection triggers a propagating torsional wave \citep{Torok2009,Pontin2011}.

The current paper addresses new observations featuring a similar magnetic topology to a fan-spine configuration. We used combined observations of three instruments onboard the Solar Orbiter \citep{Marirrodriga2021} made by the High Resolution Imager in the Extreme Ultraviolet telescope (HRIEUV) of the Extreme Ultraviolet instrument \citep{Rochus2020}, PHI \citep{Solanki2020} and SPICE \citep{SPICE2020}. The observed phenomenon is, therefore, tentatively interpreted as a propagating Alfvén wave. 
The details of the observations are given in the Section~\ref{sec:obs}. Section~\ref{analysis} is devoted to the analysis of the observations. In Section~\ref{sec:discussion} we summarise and discuss the results of the findings.

\section{Observations} \label{sec:obs}

\subsection{HRIEUV data}
\label{subsec:hri}
The images were obtained on 2022 October 18 from 14:00:00 to 14:29:55, with a temporal cadence of 5 s. In current analysis we used the data published with Data Release 5 \citep{euidatarelease5}. These observations are part of the 
R\textunderscore SMALL\textunderscore MRES\textunderscore MCAD\textunderscore AR\textunderscore Long\textunderscore Term SOOP that tracked the nearby active region (AR) for 10 days\footnote{\url{https://s2e2.cosmos.esa.int/confluence/display/SOSP/R_SMALL_MRES_MCAD_AR-Long-Term}}. At that time Solar Orbiter was close to the perihelion and its distance to the Sun constituted 0.32 au. Therefore, a pixel corresponds to $100 \times 100 \:\mathrm{km^2}$ on the Sun. 
There is no shared data with SDO AIA since the separation angle between the Earth and the Solar Orbiter was 77.6 degrees and the target AR emerged on the solar limb as seen from Earth only on October 20. 
However, there is data from the PHI and SPICE instruments on board Solar Orbiter with the field of view (FoV) including the feature that was analysed. Fig. \ref{fig:fov} shows an overview of the instrument's field of views as seen from the Solar Orbiter FSI (Full Sun Imager). The FoVs cover the targeted active region and a feature of interest for the current paper.

As a preparation step for the analysis, the images taken at different times were co-aligned using the technique described by \cite{Chitta2022} with cross-correlation as a core algorithm.

Fig. \ref{fig:video1} shows the first frame of the sequence of the images analysed in the current study. The right panel of the figure shows the first frame of the running difference image.

The feature that was used for the current analysis was located in the northern hemisphere of the Sun. As is visible from the movie (attached as electronic material), the structure is very dynamic. It consists of the footpoints connected through short loops and a spine with the length of 30 \mbox{Mm} originating from one of the footpoints. There is a remote footpoint located at around [-1440",670"] (in helioprojective-Cartesian coordinated as seen from Solar Orbiter, denoted by red circle in Fig. \ref{fig:video1}), where the spine is rooted back. 
Moreover, there is an elongated absorption feature or region with the lack of emission that is located to the left of the spine.  
\begin{figure}
  \resizebox{\hsize}{!}{\includegraphics{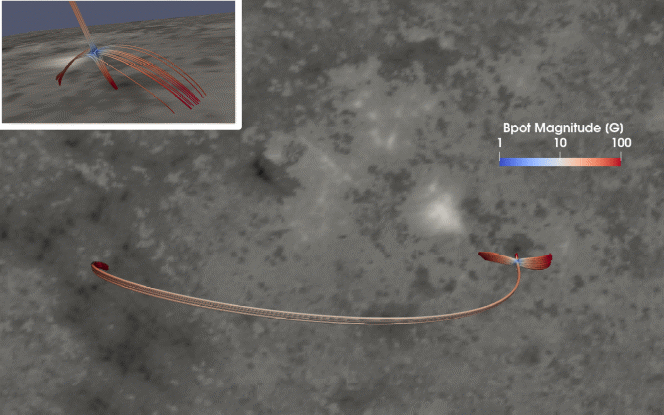}}
\caption{Field line rendering of the potential field in the vicinity of the null point. The background image represents the radial magnetic field saturated between -800 and 800 G; field lines are colored with the potential field magnitude, in logarithmic scale. The top-left inset shows an enlarged side-view of the null region.
\label{fig:extrap}}
\end{figure}

\begin{figure*}
 \centering
\begin{tabular}{cc}
\includegraphics[width=.57\linewidth]{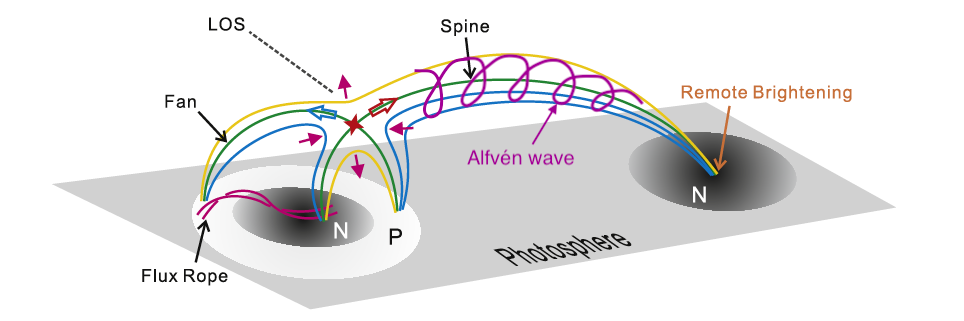} & \includegraphics[width=.35\linewidth]{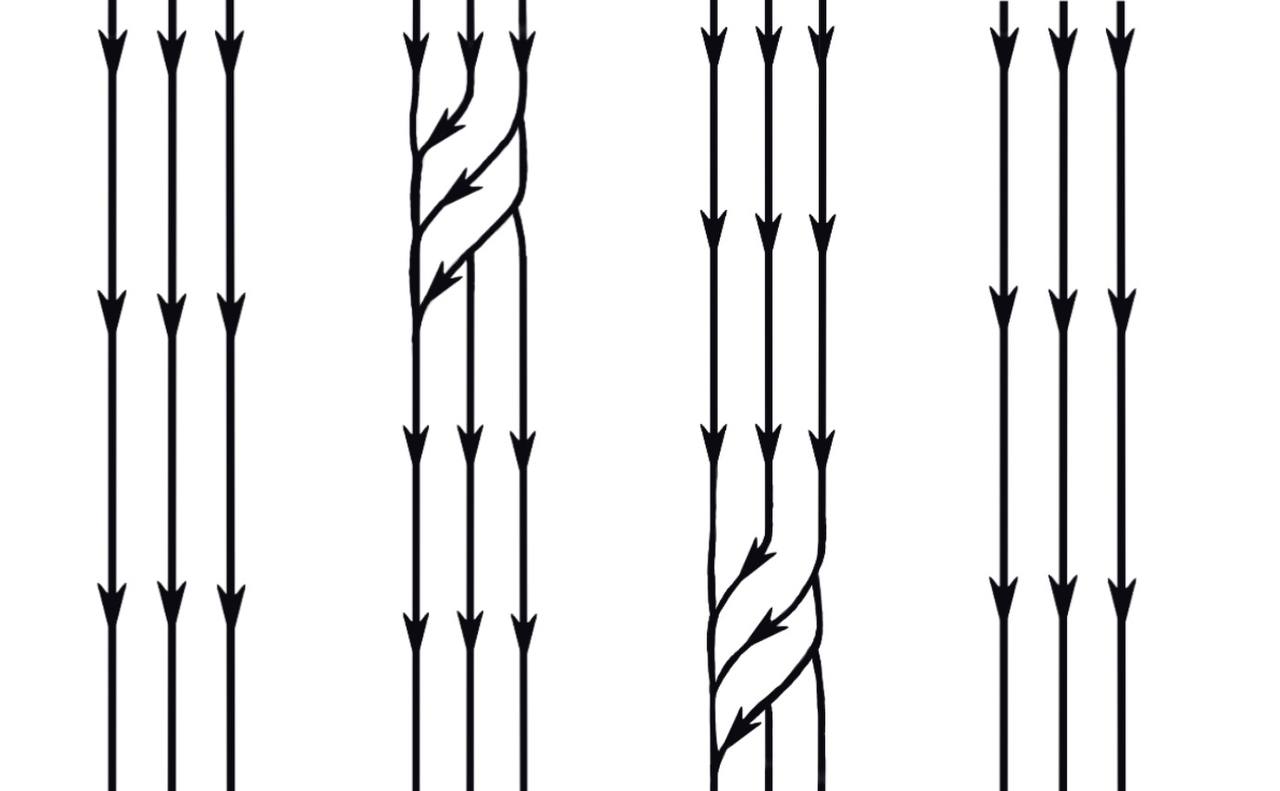} \\

(a) & (b) \\

\end{tabular}
\caption{Panel (a) shows the sketch of magnetic structure. Courtesy of \cite{Yang2020}. The purple spiral shows propagation of an Alfvén wave. Panel (b) illustrates propagation of one single Alfvén wave pulse. 
}
\label{fig:sketch}
\end{figure*}

The 174 $\AA$ HRIEUV band is primarily centered around the coronal line Fe X (core temperature around 1 MK), yet it is wide enough to also encompass multiple transition region lines. 

Though there are only 30 min of the dynamics depicted with HRIEUV, SPICE observations show that the spine exists at least for several hours. 
From both movies - original and difference, there is clear evidence of propagating material. However, apart from the propagation, there are also signatures of untwisting motion that are most clearly visible in the running difference movies. The focus of the current analysis is the torsional motions that are observed in this spine region.

\begin{figure*}
 \centering
\begin{tabular}{c}
\includegraphics[width=.58\linewidth]{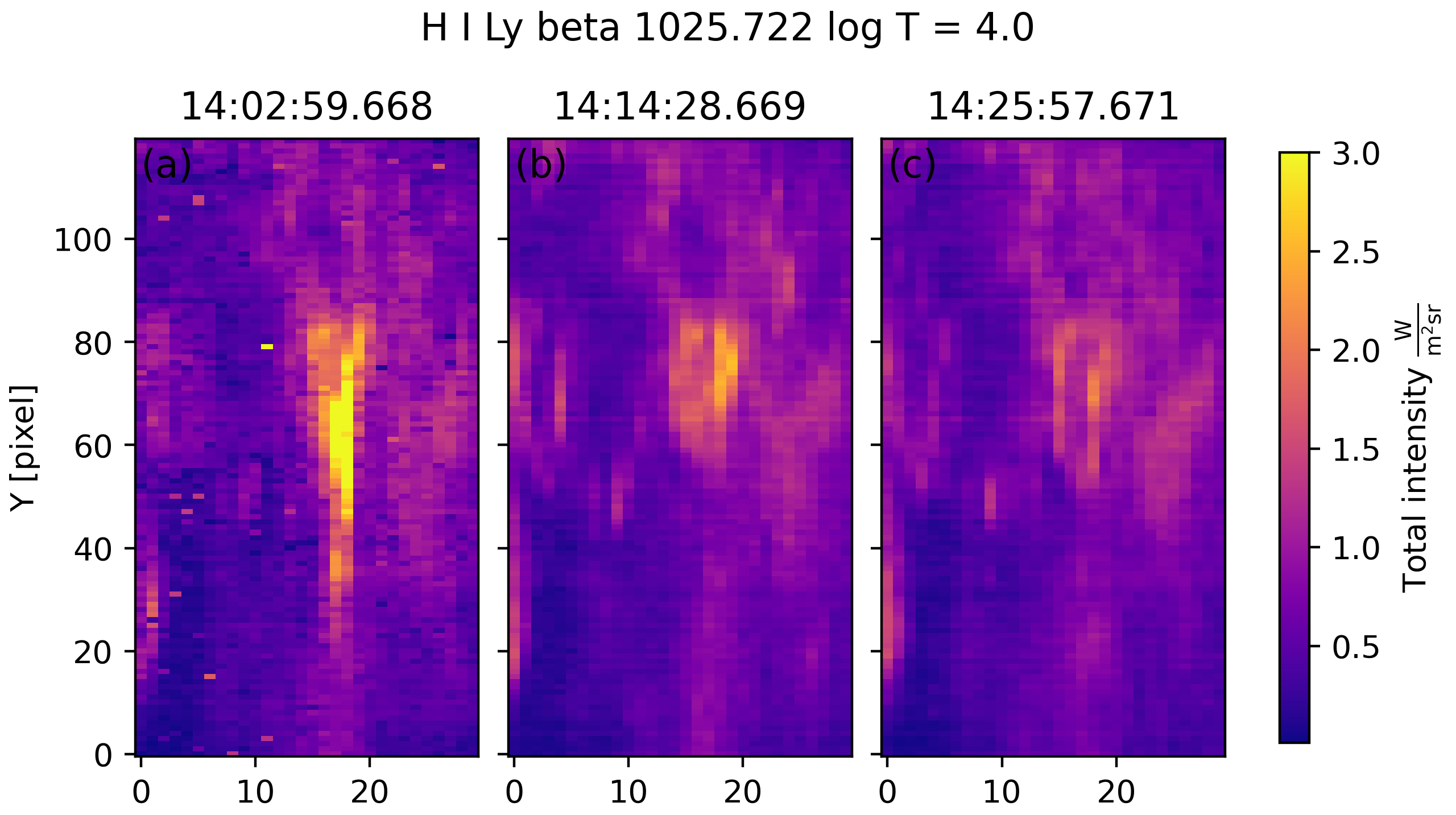}\\
\includegraphics[width=.58\linewidth]{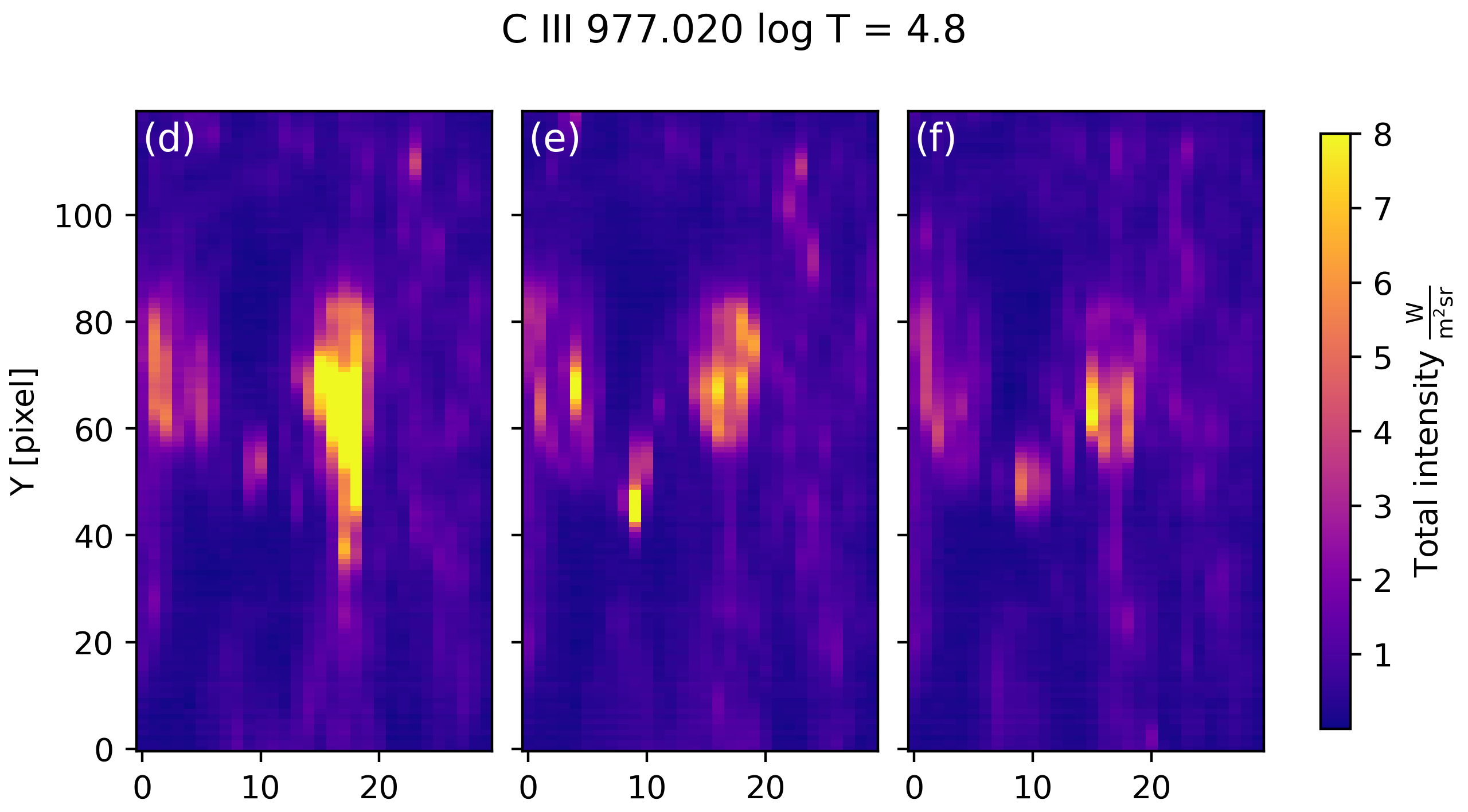}\\
\includegraphics[width=.58\linewidth]{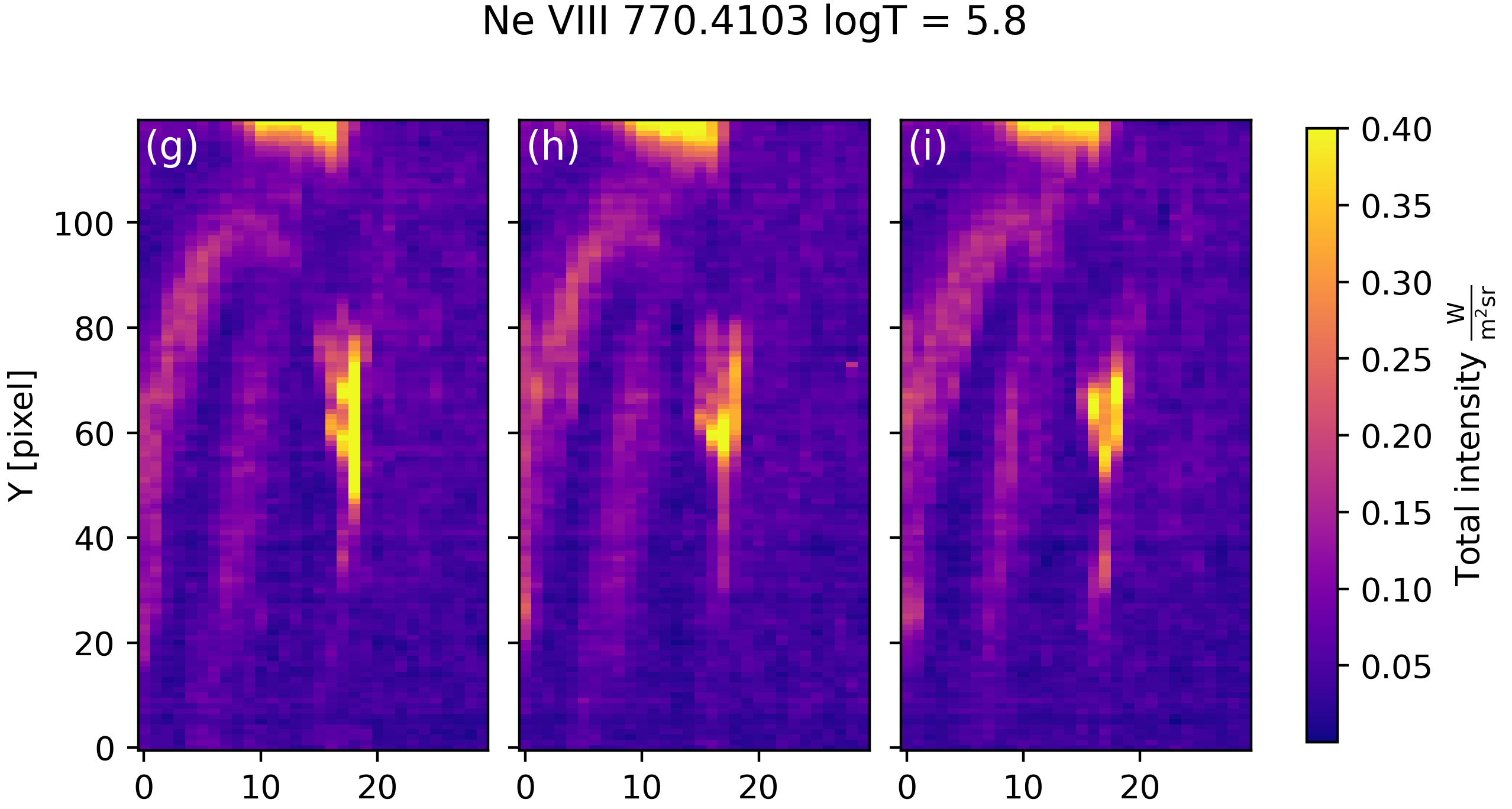}\\
\includegraphics[width=.58\linewidth]{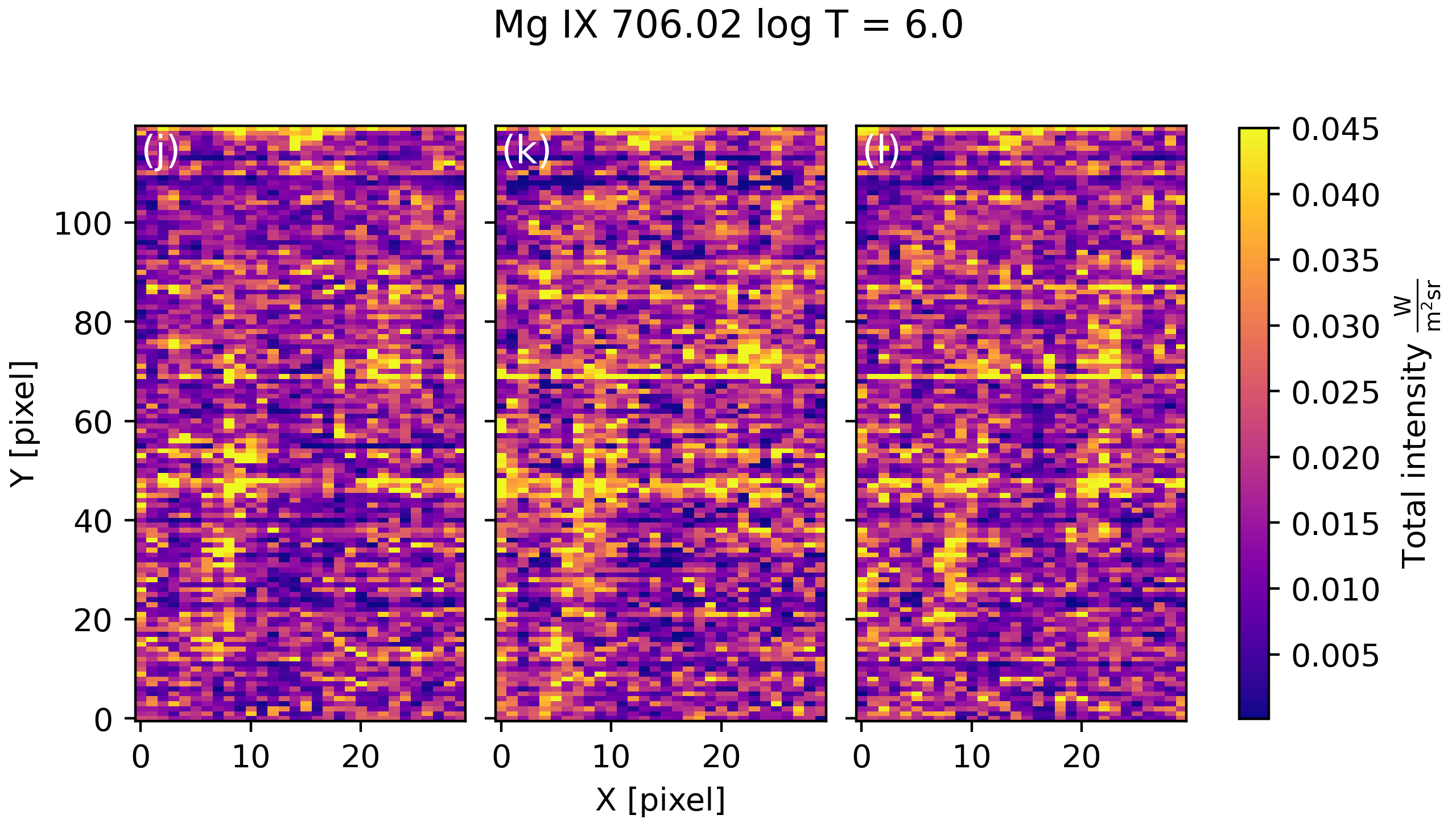}
\end{tabular}
\caption{Intensity maps obtained with SPICE for four different lines starting from the cool chromosphere and incrementally increasing up to the coronal temperatures. The first row shows three consecutive frames (14:02:59,14:14:28 and 14:25:57) for the H I Ly $\beta$ line, the second row shows intensity maps for the C III line, and the two last rows show intensity maps for the Ne VIII and Mg IX lines correspondingly. 
}
\label{fig:spice_intensity}
\end{figure*}

\begin{figure*}
  \resizebox{\hsize}{!}{\includegraphics{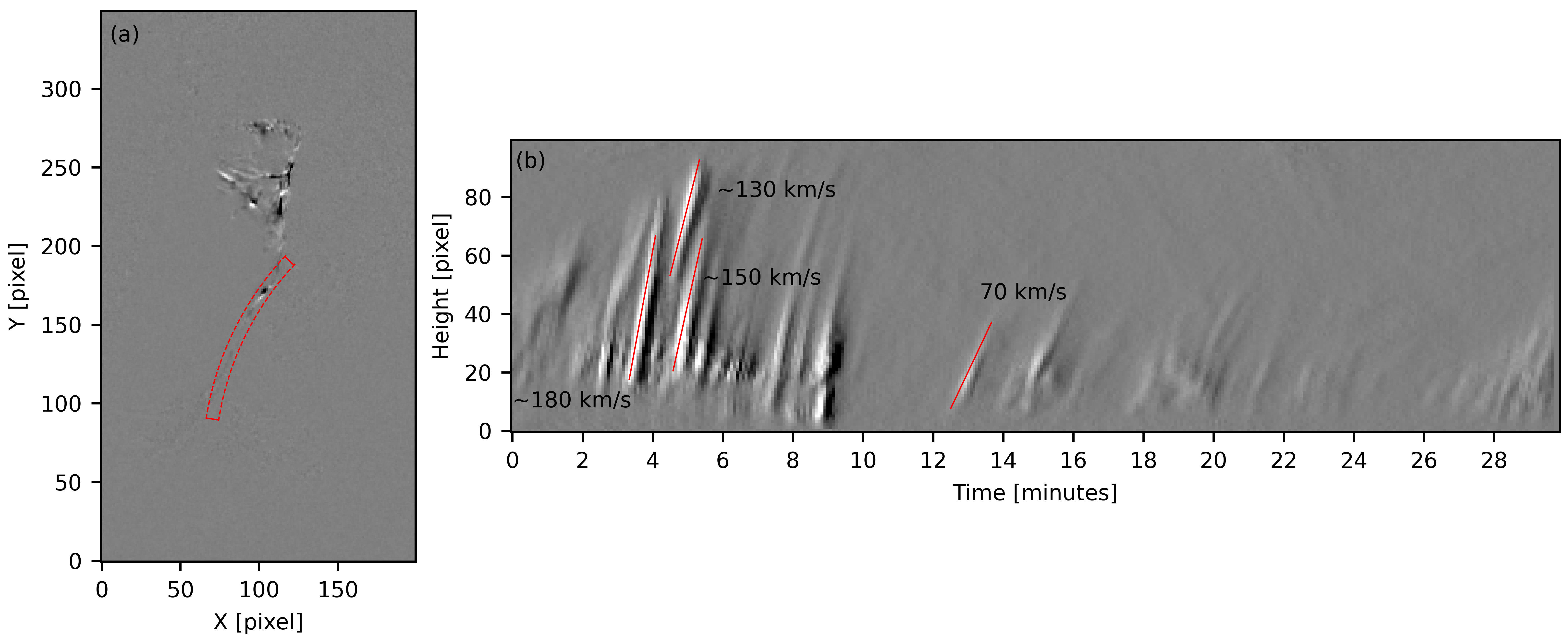}}
  \caption{Panel (a) shows the snapshot of the difference movie with the highlighted region (red dashed curves) that is used to create a time-distance map. Panel (b) shows the time distance map where propagating material is clearly visible. Red lines represent fits to the propagating wave fronts with their annotated values of propagation velocities.}
  \label{fig:propagation}
\end{figure*}

\subsection{PHI data}
\label{subsec:phi}

Data acquired by the PHI High-Resolution Telescope (HRT) \citep{Gandorfer2018} is used to get an insight into the magnetic configuration of the feature. There is one magnetogram obtained at 14:15:03 with a resolution of 0.5" per pixel, publicly available from the SOAR archive\footnote{See \url{https://www.mps.mpg.de/solar-physics/solar-orbiter-phi/data-releases} for more information about the employed PHI data release} . As a preprocessing step, the images of the HRT and EUI were co-aligned manually by visual inspection such that the fan region located to the north side of the spine and active region loops (shown by red dashed curves in Fig. \ref{fig:PHI_full} (a)) on the west of the feature would match regions with magnetic field enhancements. Active region loops originate from the sunspot shown by black circle. 

Fig. \ref{fig:PHI_full} (a) shows the cropped field of view of both telescopes aligned with each other. From the comparison we judge the alignment to be sufficiently accurate for the purposes of this work.

Fig. \ref{fig:PHI_full} (b) and (c) show that there are bipolar magnetic field concentrations corresponding to the short loops footpoints at the base of the spine. Apart from that,  there is a strong negative magnetic field corresponding to the location where the spine is rooted. And for the remote footpoint of the spine, there is a magnetic field of opposite polarity. Therefore, the magnetic configuration, in this case, resembles the fan-spine configuration described in \citet{Yang2020} as shown in Fig. \ref{fig:sketch} (a).

A potential field extrapolation shown in Fig. \ref{fig:extrap} confirmed the presence of a coronal null point and, by that, of the fan-spine topology. The spine in the extrapolation is however not oriented in the north-south direction (as in the observation), likely due to lack of information of the magnetic field on large scale, outside of the PHI field of view. However, we use the extrapolation to gain approximate information about the scaling of the magnetic field along the spine, which must be taken as order of magnitude estimate only. The spine starts from 177 G in the photosphere to zero in the null. As it keeps raising in height, reaches 13 G at the spine top. The spine then bends down and correspondingly, the magnetic field strength increases up to 220 G at the photospheric footpoint. The purple spiral in Fig. \ref{fig:sketch} (a) shows the propagation of torsional Alfvén wave. The Alfvén wave in this case looks like a propagating twist that is shown in Fig. \ref{fig:sketch} (b).

\begin{figure*}
 \centering
\begin{tabular}{ccc}
\includegraphics[width=.32\linewidth]{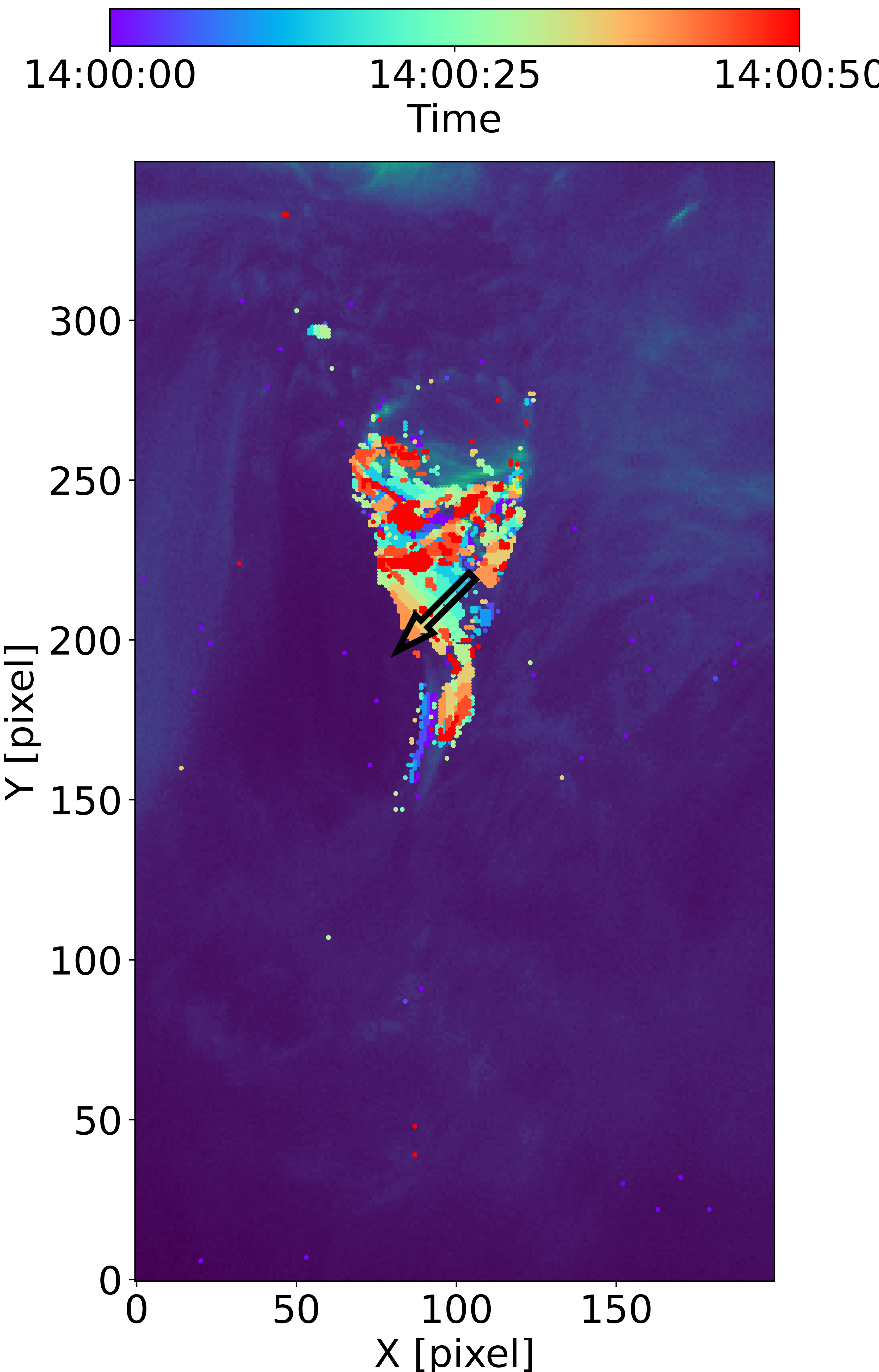} & \includegraphics[width=.32\linewidth]{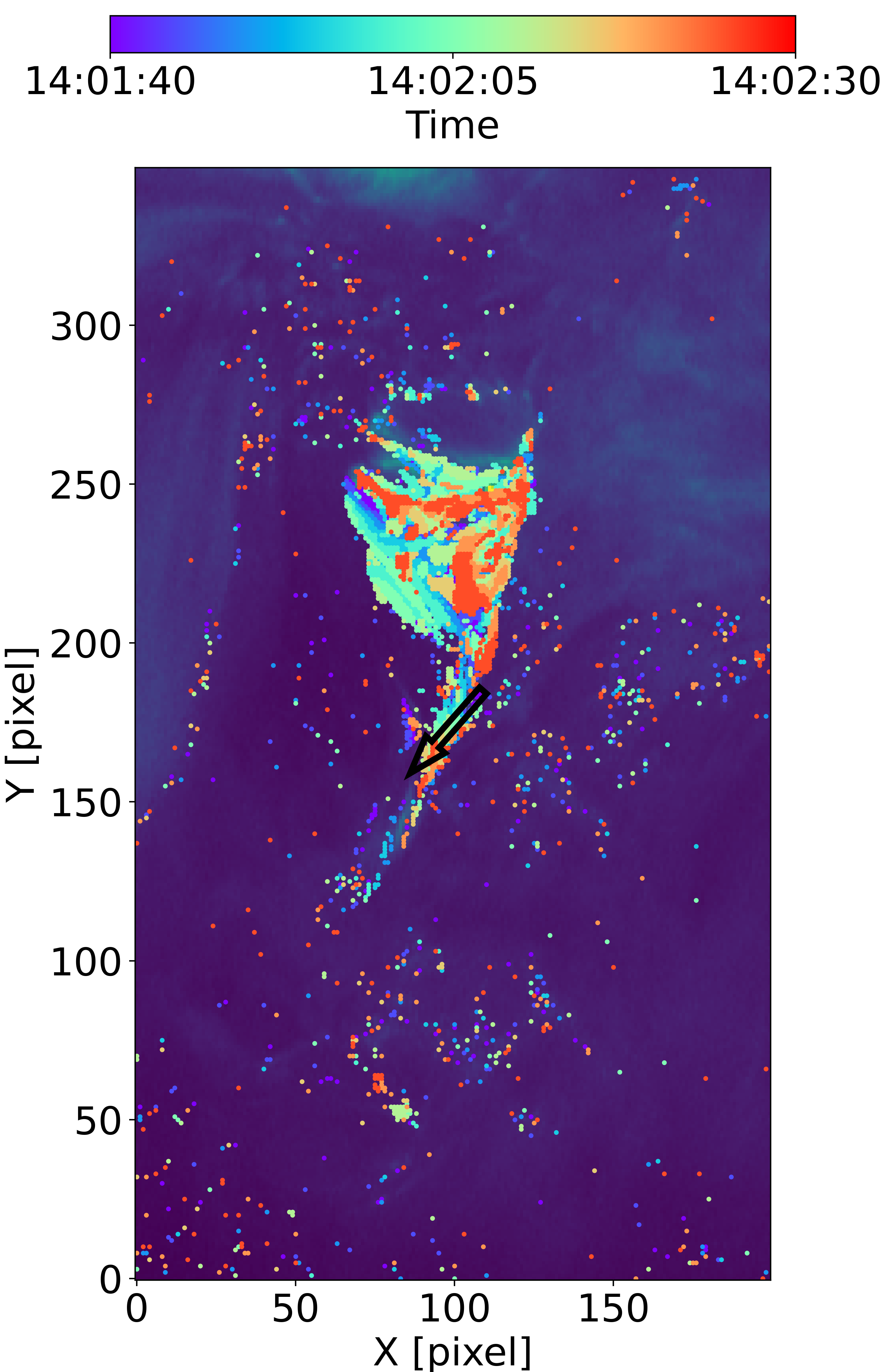} & \includegraphics[width=.32\linewidth]{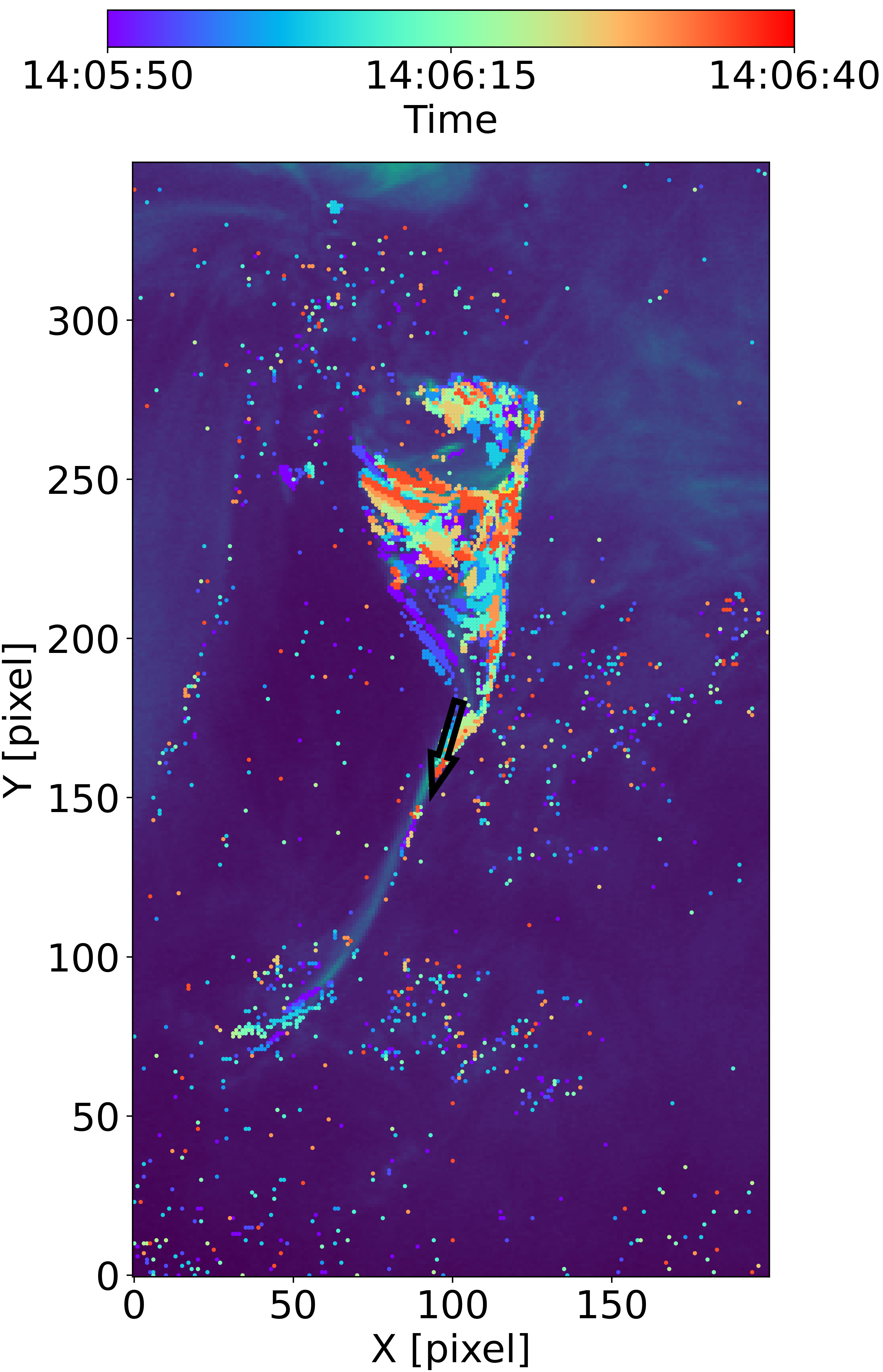}\\

(a) & (b) & (c)   \\

\end{tabular}
\caption{Snapshots of running ratios plotted on top of the HRIEUV images. The white arrows represent the direction of plasma material movement. The color bar represents intervals of time used to construct the running ratio. 
}
\label{fig:ratio}
\end{figure*}

\subsection{SPICE data}
\label{subsec:spice}
SPICE ran a dynamic sequence, rastering with the 4" slit and step 6" to give a field of view of 766"$\times$768", with 5 s exposure time. While the pixel size is 4"$\times$1", the spatial resolution is dominated by point spread function (PSF) which will be discussed in the Section \ref{subsec:doppler}. 
For the present analysis we used four spectral windows which include the following lines profiles: H I Ly$\beta$ 1025.72 \AA \:(log $T_e$ = 4.0 K), C III 977.03 \AA \:(log $T_e$ = 4.8 K), Ne VIII 770.42 \AA \:(log $T_e$= 5.8 K ) and Mg IX 706.02 \AA  \:(log $T_e$ = 6.0 K). 
This selection is the most representative of different layers of the solar atmosphere. 

We use L2 data (data release 3\footnote{\url{https://doi.org/10.48326/idoc.medoc.spice.3.0}}), which are corrected for instrumental artefacts and radiometrically calibrated, to create intensity and Doppler maps through a Gaussian fitting of the spectral lines. 

As the cadence of SPICE raster is 12 minutes, there are three raster images obtained for the whole period of the feature observed by HRIEUV. They are shown in Fig. \ref{fig:spice_intensity} where each row corresponds to a different spectral line. The Fig. \ref{fig:spice_intensity} shows the multi-thermal nature of the structure, with, at each time (a given raster position), brightenings dominating in different areas for the three temperatures. 

\begin{figure*}[h!] 
\centering
\includegraphics[width=.9\linewidth]{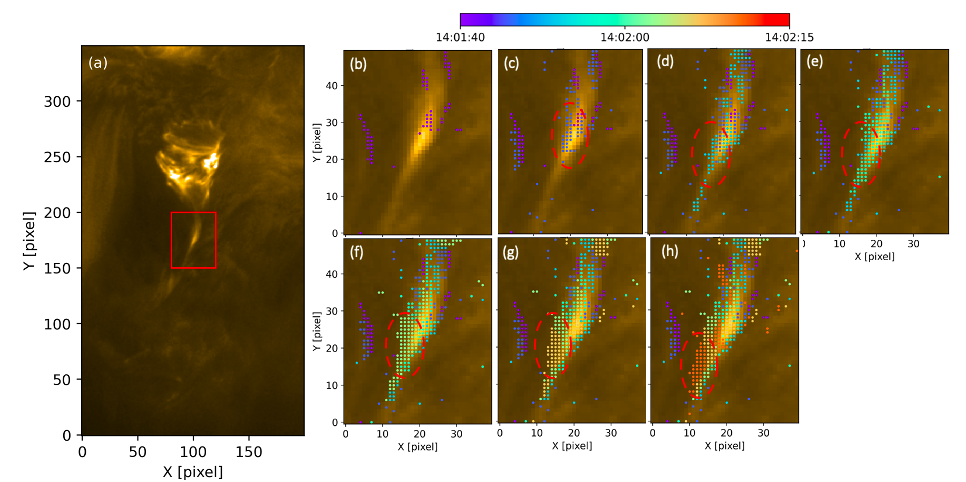}\\
\caption{Snapshots of running ratios. Panel (a) shows the spine, where the red rectangle highlights the area of interest. Panels (b)-(h) display zoomed-in views corresponding to different time frames starting from 14:01:40 and separated by 5 seconds. Points are colored according to the time, with the color bar shown at the top.}
\label{fig:pulse}
\end{figure*}

Ne VIII spectral line's depicted temperature is the closest to the HRIEUV 174 $\AA$ channel's peak temperature and shown in Fig. \ref{fig:spice_intensity} (g), (h) and (i) panels. Therefore, there we see intensity enhancements having approximately the same shapes - the spine and two bundles of loops on the left side. Over the 30 minutes of recording, intensity shows local enhancement and dimming, confirming the short time scale dynamic seen in HRIEUV while remaining about constant as a whole. 
In cooler lines like H I Ly $\beta$ (panels (a), (b) and (c)) and C III (panels (d), (e) and (f)) in the first frame there is an intensity enhancement following the whole feature. As for the consecutive frames, the intensity enhancement is much weaker in the spine region, and the fan region appears dimmer.  

The C III line shows the highest intensity enhancement hinting at the fact that the spine has a counterpart at chromospheric temperature.
However, when it comes to the hotter line like Mg IX (shown by panels (j), (k) and (l) in Fig. \ref{fig:spice_intensity}) the feature is not visible at all there, pointing out the fact that the plasma temperatures at this region correspond to chromospheric and low transition region temperatures.

\section{Data analysis}\label{analysis}

In order to estimate the propagation speed of the material, time-distance maps based on HRIEUV were constructed. 
The region for the time-distance map is indicated in Fig. \ref{fig:propagation} (a) with red dashed lines and it is chosen such that it encompasses the whole spine region in the transverse direction. 
The expected Alfvén speed $V_{A} = B/ \sqrt{\mu_0 \rho}$ is around $180 \:\mbox{km/s}$, assuming the plasma density of the spine material to be $10^{-10}\: \mathrm{kg \cdot m^{-3}}$ as in \citet{Kohutova2020} and taking magnetic field strength of 20 G as derived from extrapolation. The density assumption is based on data from SPICE that confirms that the plasma comprising the spine exhibits transition region and chromospheric temperatures. Since the uncertainty in the Alfvén speed depends solely on this density assumption, its error is half that of the density error. In the current case the measured propagation velocities are indicated in Fig. \ref{fig:propagation} (b) and are consistent with the expected Alfvén speed with values of 130 km/s to 180 km/s for the beginning of the sequence \citep{Kohutova2020}.

Apart from the fact that the movie shown by Fig. \ref{fig:video1} shows untwisting propagating motion, as another qualitative evidence of the presence of rotational motions we can use the same technique as used by \cite{Long2023}. The idea is to construct series of running ratios where values of intensities corresponding to one time frame are divided by the intensities corresponding to previous time frame. This way propagation of the pixels with higher intensity can be tracked. 
The obtained running ratios are shown in Fig. \ref{fig:ratio}. 
The limit of pixels that were plotted was identified empirically via testing different cutoff values and are different for each time range shown, varying from 1.18 to 1.3. Points of blue and purple colors correspond to the beginning of the sequence and yellow and red colors to the end of the sequence. While there might be blue and red colors present on the opposite edges of the spine, it does not necessarily mean that this is evidence of a torsional motion. However, if there is a region, where pixel's color changes smoothly from blue to red throughout the whole color sequence of the color bar, that would indicate that we are following the same plasma blob through different time frames. Therefore, if this motion is directed not along the spine, but across it, this can be considered as a signature of side-to-side or rotational motion. 


While this is only a qualitative indication of the rotation motions, one of the quantitative ways is to construct transverse slits and analyse the produced time-distance maps for movement across the axis of the spine. 

To illustrate the localization of the wave pulse, images were generated using the same technique, plotting color-coded points obtained through the calculation of running ratios (Fig. \ref{fig:pulse}). Here, the presentation is frame by frame, with each frame separated by 5 seconds, facilitating the tracing of the pulse's propagation. Panels (b)-(h) depict a zoomed-in view of the red rectangle in panel (a). The red dashed ellipse highlights the pulse being tracked. It can be observed that as time progresses through panels (b) and (c), the pulse moves toward the west and south. 
As depicted in the figure, the pulse is localized, as shown by the sequence of figures, approximately 10$\times$20 HRIEUV pixels, which is much smaller than the spine itself.

\begin{figure*}[h!] 
\centering
\includegraphics[width=0.8
\textwidth]{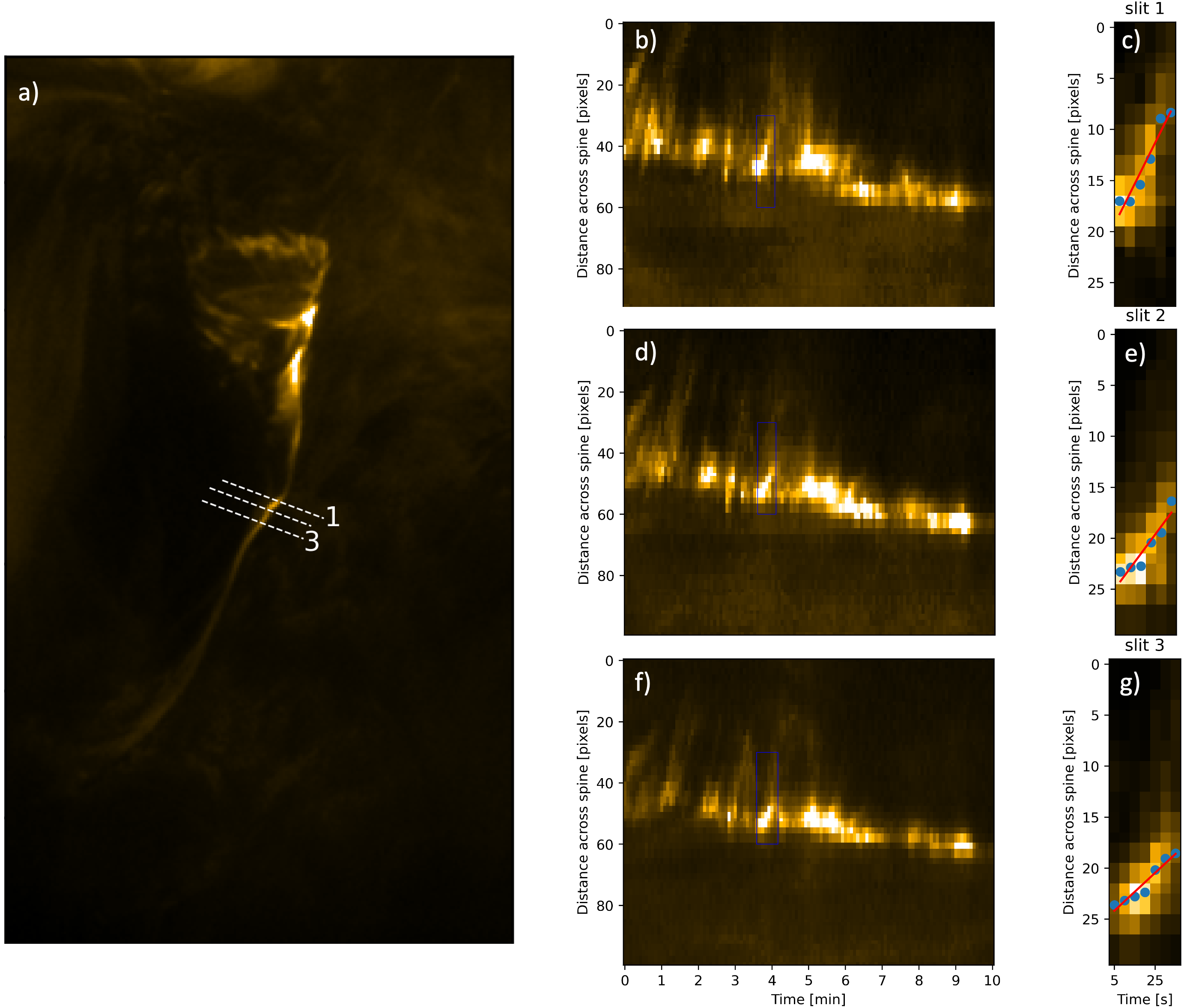}
\caption{Slits that were used to construct time-distance maps are shown by white dashed lines (left column). Numbers denote the number of each slit. Time-distance maps constructed from the slits taken across the spine (middle column) to show the evidence of the movement in transverse direction. The third column shows a zoomed view with the fitted points to the propagating pulse. The zoomed view is indicated by a blue rectangle in each figure in the middle column. }
\label{fig:transverse}
\end{figure*}

Fig. \ref{fig:transverse} shows the time-distance maps constructed from the slits that are indicated on Fig. \ref{fig:transverse} (a) by white dashed lines. In order to measure the propagation speed of the plasma, in each frame the intensity was fitted with a Gaussian function across, and the obtained centroids were fitted with a line (as denoted in Fig. \ref{fig:transverse} (c), (e) and (g)). The determined angle of the line is translated into physical speeds which constitute 26 \mbox{km/s}  for slit 1, 37.7 \mbox{km/s} for slit 2 and 56.3 \mbox{km/s} for slit 3. 

The obtained values of the rotational velocities can hint to what Doppler velocities to expect from the spectral measurements from SPICE. 

Moreover, one has to note the amplitudes of the observed waves. On average, they constitute 20 - 30\% of the local Alfvén speed indicating that linear regime assumptions cannot be applied in this case and we enter a nonlinear regime. 

\subsection{Doppler maps}
\label{subsec:doppler}
In this section we analyse Doppler maps to find line-of-sight (LoS) evidence of torsional waves. 

According to \cite{Plowman2022}, the PSF of SPICE is elongated in spectral($\lambda$) and slit-aligned ($y$) directions and tilted by approximately 15 degrees in the ($y - \lambda$) plane. The tilt in the spatial-spectral plane results in the creation of an illusionary dim feature next to a bright feature. That results in the creation of artificial signals in Doppler maps following intensity gradients. 

\begin{figure*}[h!] 
\centering
\includegraphics[width=\textwidth]{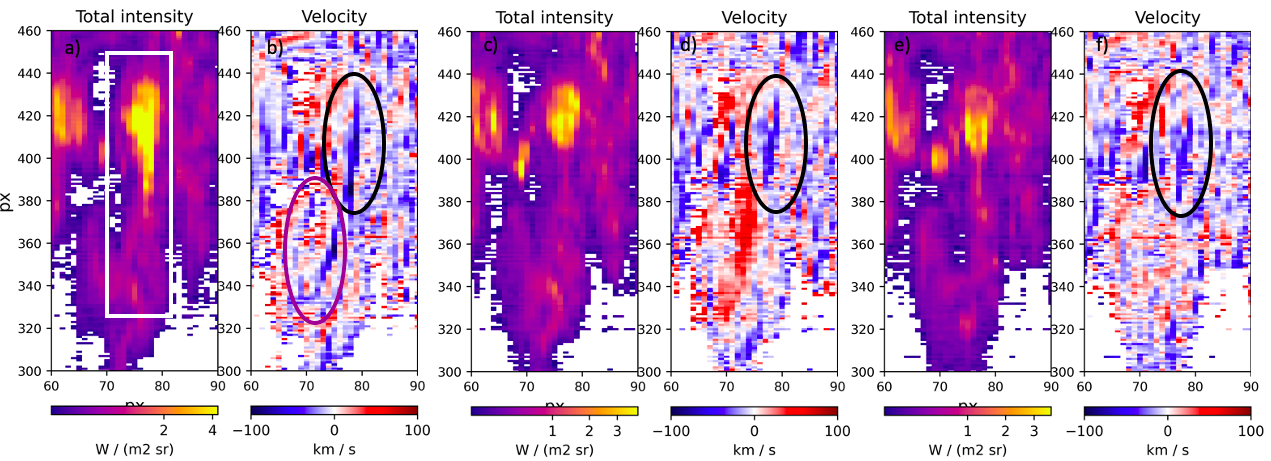}
\caption{Intensity and Doppler maps obtained with the SPICE C III line for three time frames (14:02:59,14:14:28 and 14:25:57): (a) and (b) panels show maps for the first time frame, (c) and (d) panels for the second, and (e) and (f) panels for the last frame. White rectangle in panel (a) shows the location of the spine. Black ellipses in panels (b), (d) and (f) show location with the strong signal that matches high intensity gradient in panel (a). Magenta ellipse in panel (b) shows location with strong Doppler signal that does not correspond to the high intensity gradient.}
\label{fig:SPICE_doppler}
\end{figure*}

Fig. \ref{fig:SPICE_doppler} shows the C III intensity and Doppler maps obtained by SPICE for the three different times. Pixels  that have a very low signal in intensity maps and zero velocity in Doppler maps appear as white. The location of the spine is denoted by the white rectangle in Fig. \ref{fig:SPICE_doppler} (a). There are two distinct locations to note in Doppler maps - region that corresponds to the high intensity gradient (shown by the black ellipse) and the one that does not (shown by the magenta ellipse in panel (b)). Across all time frames, Doppler maps consistently display a pronounced strong blue shift in the region marked by the black ellipse. However, the area identified by the magenta ellipse only exhibits a discernible signal, distinguishable from noise, in panel (b). Moving to the next time frame, as depicted in panel (d), strong red shifts appear to the left of the spine, possibly associated with an absorption feature, while it's hard to say if the signal is physical of artificial. Panel (f) shows some level of red shifts but it's hard to make any conclusions about its nature. 

While it is known that the described earlier PSF issues of SPICE can create artificial signal in Doppler maps, torsional waves as well produce Doppler shifts as it is one of the primary detection features for this phenomenon. Therefore, most likely, the maps obtained by SPICE have a combination of real and artificial signals and it is important to be able to separate one from another. In order to estimate the influence of the tilted SPICE PSF, synthetic raster spectra were made using as an input HRIEUV 174 $\AA$ data. 

\begin{figure*}[h!] 
\centering
\includegraphics[width=.85\linewidth]{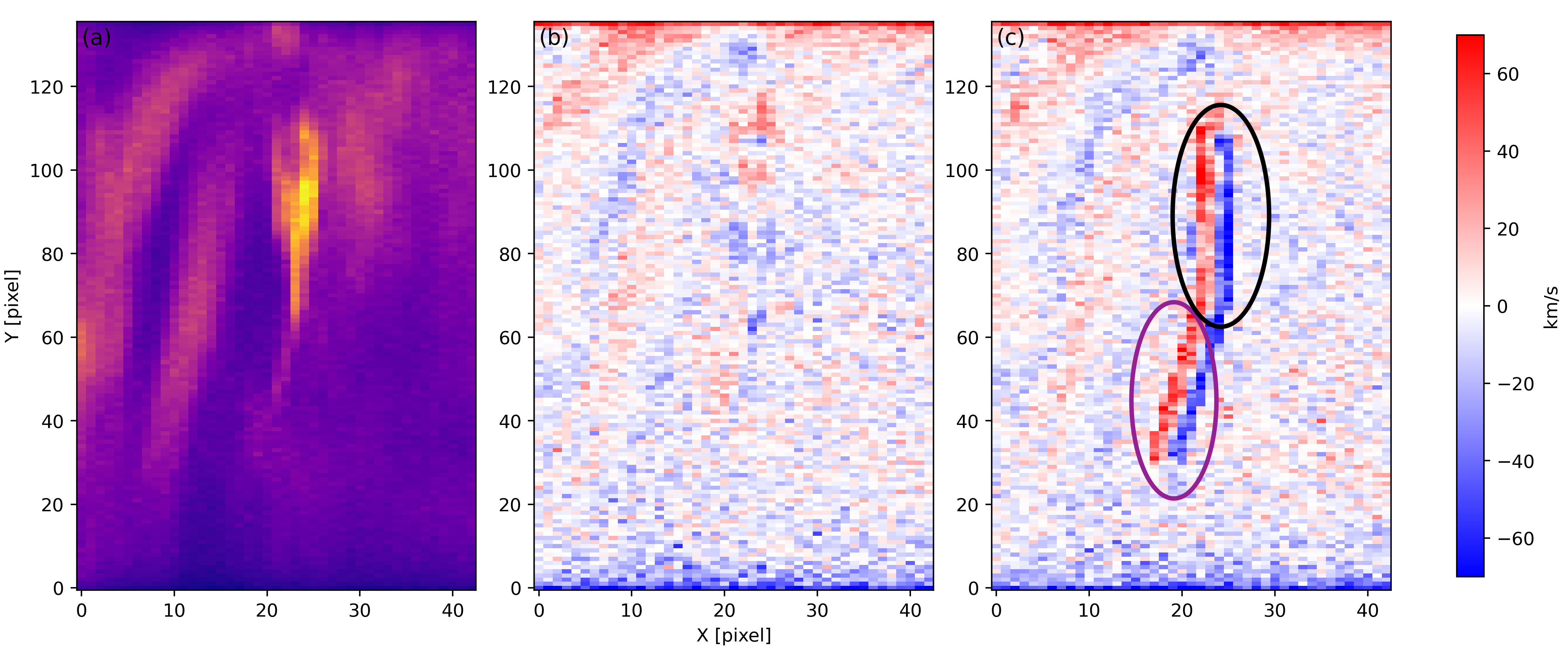}\\
\caption{Synthesized spectral rasters obtained using forward-modelled analysis. Panel (a) shows the synthesized intensity, while the (b) and (c) panels show Doppler maps (mimicking the SPICE CIII line) with and without local velocities imposed on top of the feature. The black ellipse in panel (c) shows the location of strong signal that corresponds to the high intensity gradient. The magenta ellipse outlines the location that does not correspond to any intensity gradient. }
\label{fig:synth_doppler}
\end{figure*}

It is important to mention that the synthesised images will correspond only partially to temperatures the SPICE lines depict. Therefore, it is impossible to fully reproduce SPICE images which potentially can lead to the different behaviour shown in both intensity and Doppler maps. Although the Ne VIII line has the closest temperature coverage to the 174 $\AA$ channel, its intensity is much smaller compared to the CIII line. For this reason it makes more sense to compare the modeled intensity maps with that of the SPICE Ne VIII line. However, when it comes to Doppler maps, the Ne VIII line does not show any signal due to the mentioned weak signal issue. Therefore, the Doppler maps will be compared to those from the CIII line.

The EUI images need to be scanned as it is done by SPICE. The exposure time for SPICE is 5 seconds and slit width is $4"$ with a step of $6"$ which for HRI with a pixel resolution of $0.5"$ means that every 5 seconds a column with the width of 8 pixels in HRI is saved. However, the nonzero width of the slit means that after every 8 columns of HRI data being saved, there are 4 columns of data being lost. As a result, we have a scanned image, but still with the resolution of HRI, therefore the next step is downsampling the scanned image to the SPICE resolution. 

The next step is to supplement the data with a spectral dimension. The width of an observed emission line is defined by the thermal broadening, non-thermal motions, and instrumental broadening and can be expressed as follows in terms of the full width at half maximum (FWHM):
\begin{equation*}
    w^2 = w_\mathrm{th}^2+w_\mathrm{nth}^2+w_\mathrm{I}^2
\end{equation*}

The thermal broadening is calculated from the peak temperature $T = 0.8\:\mathrm{MK}$ of the 174 $\AA$ channel as following:

\begin{equation}
    w_{\mathrm{th}} = \sqrt{\frac{8 \ln{2} \cdot k_B T} {m_\mathrm{ion}}}\frac{\lambda_0}{c}
\end{equation}
where $k_B$ is the Boltzmann constant, $c$ is the speed of light, $\lambda_0$ is the expected centroid wavelength of the line (which is 174 $\AA$ in our case)  and $m_{ion}$ is mass of the emitting ion. The condition of $\lambda = \lambda_0$ results in zero Doppler shift. 

The non-thermal broadening $w_{nth}$ can be caused by several different processes including for example, turbulence \citep{Pontin2020}, small-scale flows \citep{DePontieu2015}, waves \citep{Chae1998}. According to \cite{Chae1998}, the coronal non-thermal velocity is equal to 20 km/s and, therefore, this value was used for the current analysis. 

However, the line width in SPICE is dominated by the instrumental width $w_{I}$. As described in \cite{Fludra2021} the FWHM (estimated for 2" slit) for the Short Wavelength (SW) band is estimated to be 8 pixels and 9 pixels for the Long Wavelngth (LW)  band. This corresponds to Doppler velocities of 200 km/s. The PSF is not completely known for the 4'' slit, so we use the properties derived for the 2'' slit. 
Now, knowing the parameters of the Gaussian and the peak intensity $I_p$ in each pixel, we can construct a wavelength dimension as it would have been observed by SPICE:
\begin{equation}
    I = I_p\cdot \mathrm{exp}[ - \frac{(\lambda-\lambda_0)^2}{2\sigma^2}]
    \label{eq:1}
\end{equation}
where the width of the Gaussian $\sigma$ is related to the FWHM $w$ through $w = 2 \sqrt{2 \ln{2}} \sigma$. 

As a result, we have a 3D cube of the data which was then convolved with the SPICE PSF in ($y-\lambda$) space as it is described in \cite{Plowman2022}.  


The last step for this simulation is to fit the intensities in the wavelength dimension with a Gaussian again to obtain Doppler shifts. The total intensity is obtained by integration of the Gaussian over a wavelength range. Poisson noise that depends on intensity such that it is comparable to noise in SPICE was added. The results (for the time frame 14:14:28) are shown in Fig. \ref{fig:synth_doppler}. Panel (a) shows intensities obtained from line fitting and panel (b) shows the Doppler map. 

However, this simulation alone is not enough to distinguish between signal caused by rotational motions and artificial PSF signal contributions in SPICE Doppler maps. For that we make another simulation in a similar fashion but now with local Doppler velocities imposed on top of the feature. The model for the velocity field is chosen to be a simple case of a rigid body rotation. Details of the velocity field are given in appendix \ref{appendix}. The result is shown in Fig. \ref{fig:synth_doppler} (c).

With that being said, the synthesized intensity map depicted by Fig. \ref{fig:synth_doppler} (a) follows the same shape as observed by the SPICE C III and Ne VIII clines shown in second and third rows of Fig. \ref{fig:spice_intensity}. Specifically, the latter reproduces the signal coming from the loops located to the left of the spine. Similarly, the signal coming from the spine weakens for lower values of the y-coordinate. 

The base of the feature is more extended (about 6 columns in the C III raster) than the dome (Ne VIII, 4 columns). The synthetic raster created using HRIEUV resembles, but is not precisely identical to the Ne VIII. The shape is something in between Ne VIII and C III. This suggest that the dynamics within the feature is associated with multi-thermal plasma evolution not completely caught by the HRIEUV. We suggest that the  torsional waves are within these multi-thermal structures.

As expected from the effects of the PSF, panel (b) with  Doppler maps without local velocities imposed shows that there is a signal following the gradients of the intensity region matching the signal from the loops and the spine. The signal in the spine region reaches a maximum of 20 km/s for both blue and red shifts. The map with the velocities imposed depicted by panel (c) understandably shows a strong signal following the central lines of the rotation reaching up to 60 km/s which is comparable to values measured by SPICE. In order to match synthetic spectra with observations, Fig. \ref{fig:synth_doppler} (c) should be compared to the Fig. \ref{fig:SPICE_doppler} (b). The signal is modulated by the effect of the PSF leading to various non-uniformities as opposed to the imposed velocity field coming from the intensity gradients. While the red shifts were there on the panel (b) of Fig. \ref{fig:SPICE_doppler}, the strong blue shifts are the new phenomenon which matches the SPICE observations indicating that this signal should be physical under the assumption that the recovered PSF is correct. 

Furthermore, a notably strong Doppler signal persists in a region marked by the magenta ellipse in Fig. \ref{fig:synth_doppler} (c), even in the absence of a locally imposed velocity. Considering that there are no intensity enhancements in the lower spine region, yet both synthesized and observed images exhibit a signal, provides substantial evidence for its authenticity under the assumption that the recovered PSF is mostly correct.

\subsection{Magnetic field perturbation}
\label{subsec:elsasser}
With the use of the Elsässer variables, one can estimate perturbated magnetic field compared to the background field. In incompressible MHD, counterpropagating waves are  described by Elsässer variables $\vec z^{\pm}=\vec V \pm \frac{\vec B}{\sqrt{\mu \rho} }$,  where  $\vec V$ is the velocity of the plasma, $\vec B $ is the magnetic field, $\mu$ and $\rho$ are the magnetic permeability and density \citep{Elsasser1950}. Positive corresponds to the outward propagating wave and negative to inward propagating wave. For our case we have $\vec z^-$ which equals to zero since we have an Alfvén wave. With the use of perturbed over equilibrium variables (denoted with index 0, with no assumption on its amplitude) and background variables, the expression can be written as follows:

\begin{equation*}   
\vec z^{\pm} = \vec z_0^{\pm}+z'^{\pm} = (\vec V_{0} \pm \frac{\vec B_0}{\sqrt{\mu \rho}} )+ ( \vec V' \pm \frac{\vec B'}{\sqrt{\mu \rho}}) 
\end{equation*}

Working with the absolute values of vectors one can deduce the perturbed magnetic field:

\begin{equation*}   
|\vec B'| = \frac{|\vec V'|\cdot|\vec B_0|}{|\vec V_0|}
\end{equation*}

Taking estimates from the observations for the velocities and magnetic field of 20 G from the extrapolation, we can conclude that the perturbed magnetic field should be around 4.5 G.

\section{Discussion and conclusion} \label{sec:discussion}

The current study presents observations of a dynamic feature consisting of a spine feature. This feature was observed by the EUI HRIEUV, PHI HRT, and SPICE instruments onboard Solar Orbiter. The HRIEUV movies give a strong impression of torsional motions in the spine structure. In this work we have confirmed these torsional motions from the base of the spine southward (Fig. \ref{fig:ratio}) and into the spine by showing the simultaneous presence of: southward longitudinal motions (Fig. \ref{fig:propagation}), transverse motions in the plane of sky (Fig. \ref{fig:transverse}), line-of-sight motion (Fig. \ref{fig:SPICE_doppler}). In addition, since these torsional motions have a speed consistent with a local estimate of the Alfvén speed, we tentatively identify these torsional motions as torsional Alfvén waves. 

There is a signature of torsional motions visible in difference and original movies. Possible evidence of that can be found in time-distance maps obtained from transverse cuts with velocities in the transverse direction constituting 26-56 \mbox{km/s}. Running ratios also point out the fact that there is indeed not only the propagation of the matter but also side-to-side motion. 
Measured propagation speed constitutes values of 130 - 180 \mbox{km/s} which is consistent with the expected values of Alfvén speed and also similar to what has been measured by \cite{Kohutova2020}.  
Moreover, the performed forward modelling serves as an additional argument to the possible interpretation of the observed phenomenon to be torsional motion as it shows that there is a Doppler signal in a region, corresponding to the spine. 
While this analysis is beneficial for the current work, it may also be used to gain an insight into the reliability of SPICE Doppler maps, and the instrument's ability to detect rotational or line-of-sight movements. 

However, the issue of artefacts appearing in Doppler maps is quite complicated, and it has recently been recognized that the PSF is not the sole source of the false signal. Additional sources, such as the 'overlappograph' \citep{Tousey1977} can create a Doppler shift when a bright feature is present on one side of the slit. This effect is weaker when the emission is less sharp which appears to be the case for the HRIEUV 174 \AA \: channel when compared to the SPICE lines. Nevertheless, reproducing this effect in current simulations poses a challenge. Consequently, efforts to untangle the artificial signal from the physical one are still ongoing, and further work is necessary to achieve this goal. 
 
One of the primary signatures of these oscillations is their substantial amplitude. The velocity in the transverse direction accounts for 30\% of the local Alfvén speed, indicating that these waves exist within a non-linear regime. This observation would be a prime example to look for non-linear effects of Alfvén waves, due to its extremely high velocity amplitude. One could look for steepening wave fronts, turbulence formation or the existence of solitons or solitary waves. 

As a result of the significant velocity amplitude of these waves, the energy budget becomes considerably high. Computed using the formula $E = \rho v^2 V_A$, it reaches magnitudes on the order of tens of $\mathrm{kW/m^2}$ although it is important to acknowledge that this estimate is susceptible to errors arising from density assumption and velocity calculations. However, it is notably greater than the calculated energy that should be transferred to the corona, as estimated by \cite{Jess2009, DePontieu2007}. This estimated energy transfer should be on par with the findings in \cite{Kohutova2020}. While \cite{Kohutova2020} does not provide a direct calculation of the energy, the input parameters used therein are comparable to those in the current manuscript, thereby yielding similar flux values.
As for the driver of those oscillations, it is still a point of question. It has been previously observed, for example, by \cite{Kohutova2020} that torsional waves can be triggered by reconnection which is happening in this case. Moreover, \citet{Pariat2009} shows that null point reconnection can release the twist in a jet. However, the fact that hotter SPICE line like Mg IX do not have any signal at all, shows that the temperature does not increase up to a point of $\mathrm{log\: T} = 6.0\: \mathrm{K}$ thus serving as an argument against reconnection as a potential wave exciter candidate. 

However, it is important to mention that there are alternative interpretations of the event such as propagating kink wave with circular polarization \citep{Magyar2022}. With the available data and no shared view from another vantage point there is no possibility to differentiate between those two. Additionally, torsional motions of the coronal structure may be considered as evidence of torsional Alfvén waves, but other interpretations, such as untwisting of a flux rope \citep{Li2015,Liu2009}, could not be ruled out. There are multiple factors to take into account when distinguishing between them. 
\begin{itemize}
    \item The accurate estimation of the Alfvén speed in the observed spine is impossible without the differential emission measure (DEM) analysis of the structure. However, as mentioned in Section~\ref{sec:obs}, there is no shared data with the SDO AIA, and the SPICE instrument onboard Solar Orbiter lacks sufficient spectral lines to conduct DEM analysis. Given these circumstances, assuming the density value is unfortunately the only option left. However, since the Alfvén speed is only influenced by the square root of the density, even with an order of magnitude difference from the actual density, the Alfvén speed would only vary by a factor of 3.
    
	\item Due to the projection effect, the torsional motion of the spine can become apparent. For example, it can be attributed to the antiparallel motion of different plasma layers along the line of sight.

	\item The SPICE time cadence is too high to capture the prominent wave dynamics. However, high cadence spectral data is not necessarily required to confirm the presence of waves, as no periodicity is expected if the driver launches pulses. The current cadence of 12 minutes is sufficient to support the conclusion that there is LoS evidence of torsional motions.
	
	\item The consideration of pulse size is also necessary. Figures \ref{fig:SPICE_doppler} and \ref{fig:synth_doppler} may suggest that the pulse size is equal to the spine's size, supporting the interpretation of untwisting of a flux rope. However, the width of the pulse in the mentioned figures is not spatially resolved and determined by the SPICE slit width and the scanning speed. As for Fig. \ref{fig:synth_doppler}, this is the case because the velocity field is imposed based on the model of rigid body rotation. Therefore, for simplicity, it is assumed that the body rotates as a whole. 
	
	 \item The observations of the torsional motions in 174 $\AA$ intensity maps and in C III Doppler maps are not necessarily attributed to one phenomenon because the line temperatures differ significantly. Though it is true, the intensity signal in NeVIII is much smaller compared to the CIII line. Consequently, the Ne VIII line does not exhibit any signal in Doppler maps that would overpower the noise level.

\end{itemize}
 
Considering the points described above, we lean more towards the interpretation of propagating torsional Alfvén waves. However, it is also worth noting that the possibility of untwisting motion cannot be entirely ruled out. 
 
\begin{acknowledgements}
Solar Orbiter is a space mission of international collaboration between ESA and NASA, operated by ESA. The EUI instrument was built by CSL, IAS, MPS, MSSL/UCL, PMOD/WRC, ROB, LCF/IO with funding from the Belgian Federal Science Policy Office (BELSPO/PRODEX PEA 4000112292); the Centre National d’Etudes Spatiales (CNES); the UK Space Agency (UKSA); the Bundesministerium für Wirtschaft und Energie (BMWi) through the Deutsches Zentrum für Luft- und Raumfahrt (DLR); and the Swiss Space Office (SSO).
\newline

We are grateful to the ESA SOC and MOC teams for their support. 
The German contribution to SO/PHI is funded by the BMWi through DLR and by MPG central funds. The Spanish contribution is funded by AEI/MCIN/10.13039/501100011033/ and European Union “NextGenerationEU”/PRTR” (RTI2018-096886-C5,  PID2021-125325OB-C5,  PCI2022-135009-2, PCI2022-135029-2) and ERDF “A way of making Europe”; “Center of Excellence Severo Ochoa” awards to IAA-CSIC (SEV-2017-0709, CEX2021-001131-S); and a Ramón y Cajal fellowship awarded to DOS. The French contribution is funded by CNES. 
\newline
The development of SPICE has been funded by ESA member states and ESA. It was built and is operated by a multi-national consortium of research institutes supported by their respective funding agencies: STFCRAL (UKSA,hardwarelead), IAS (CNES,operationslead), GSFC (NASA), MPS (DLR), PMOD/WRC (SwissSpaceOffice), SwRI(NASA), UiO (NorwegianSpaceAgency).
\newline

TVD was supported by the C1 grant TRACEspace of Internal Funds KU Leuven. EP has benefited from the funding of the FWO Vlaanderen through a Senior Research Project (G088021N).

\end{acknowledgements}

\bibliography{References,refs_tomvd}{}
\bibliographystyle{aa}

\newpage
\appendix 
\section{Local velocity field and spectra}
\label{appendix}
\begin{figure}
  \resizebox{\hsize}{!}{\includegraphics{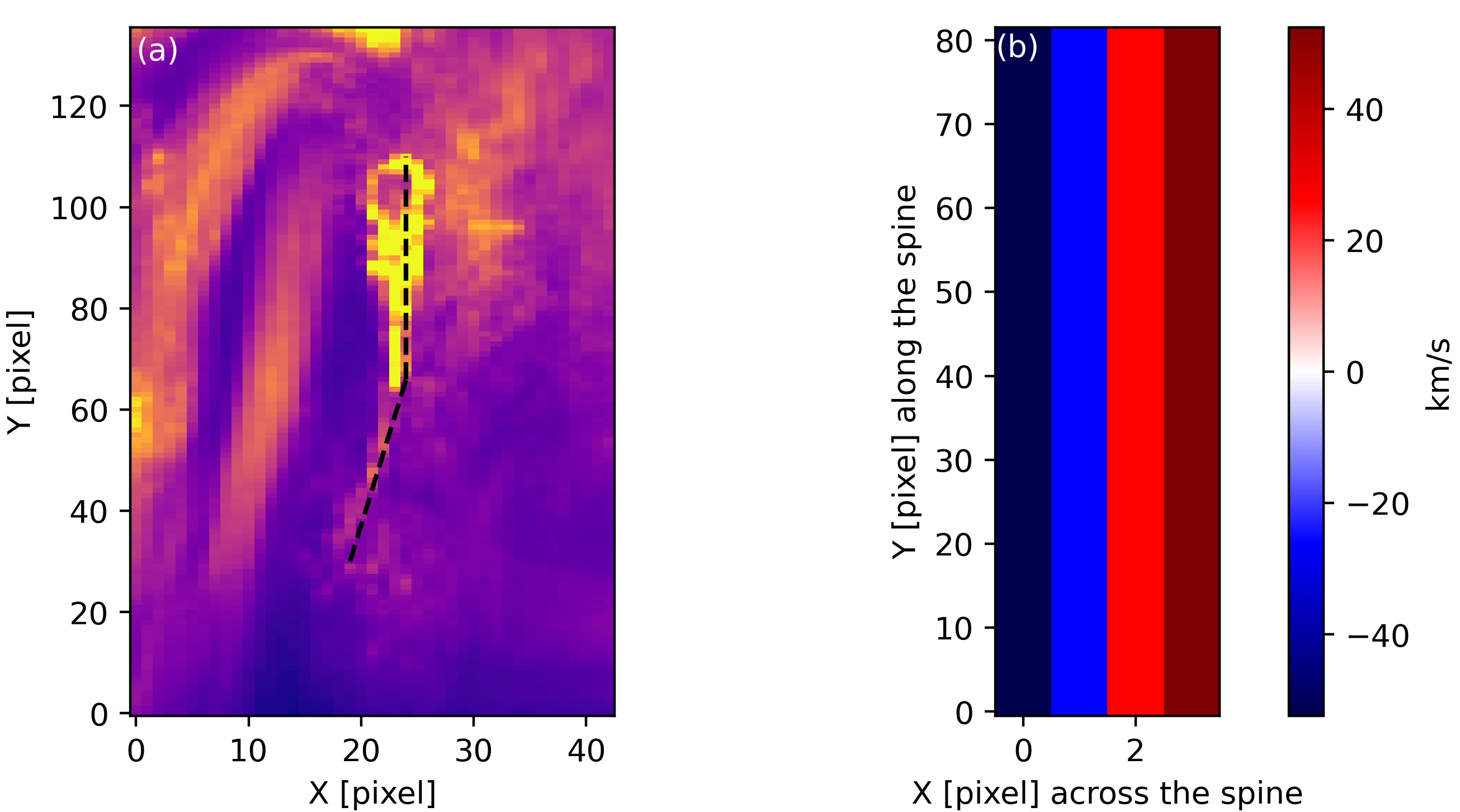}}
  \caption{Central lines of the feature identified as in rigid body rotation denoted by dashed black lines (panel (a)). Velocity field imposed on top of the feature is depicted by panel (b).}
  \label{fig:v_field}
\end{figure}

As mentioned above, local Doppler velocities were imposed on top of the feature for the forward modelling analysis. 
The model for the velocity field is a rigid body rotation. Therefore, since the current feature has a complex shape, it was fitted with two lines as shown in Fig. \ref{fig:v_field} (a) by black dashed lines. They were used as two central lines with the imposed velocity field as shown in Fig. \ref{fig:v_field} (b). Values of the velocity were taken such that they would correspond to the observed values of velocities in the transverse direction as obtained in Section \ref{analysis}. The line in the $x$ direction where negative velocities transit to positive velocities corresponds to the central line around which rotation is happening.

Then, the described velocity field was used to construct a wavelength dimension similar to Eq. \ref{eq:1}:
\begin{equation}
    I = I_p\cdot \mathrm{exp}[ - \frac{1}{2\sigma^2}((\lambda-\lambda_0(1-\frac{v}{c})))]
\end{equation}

\begin{figure}
  \resizebox{\hsize}{!}{\includegraphics{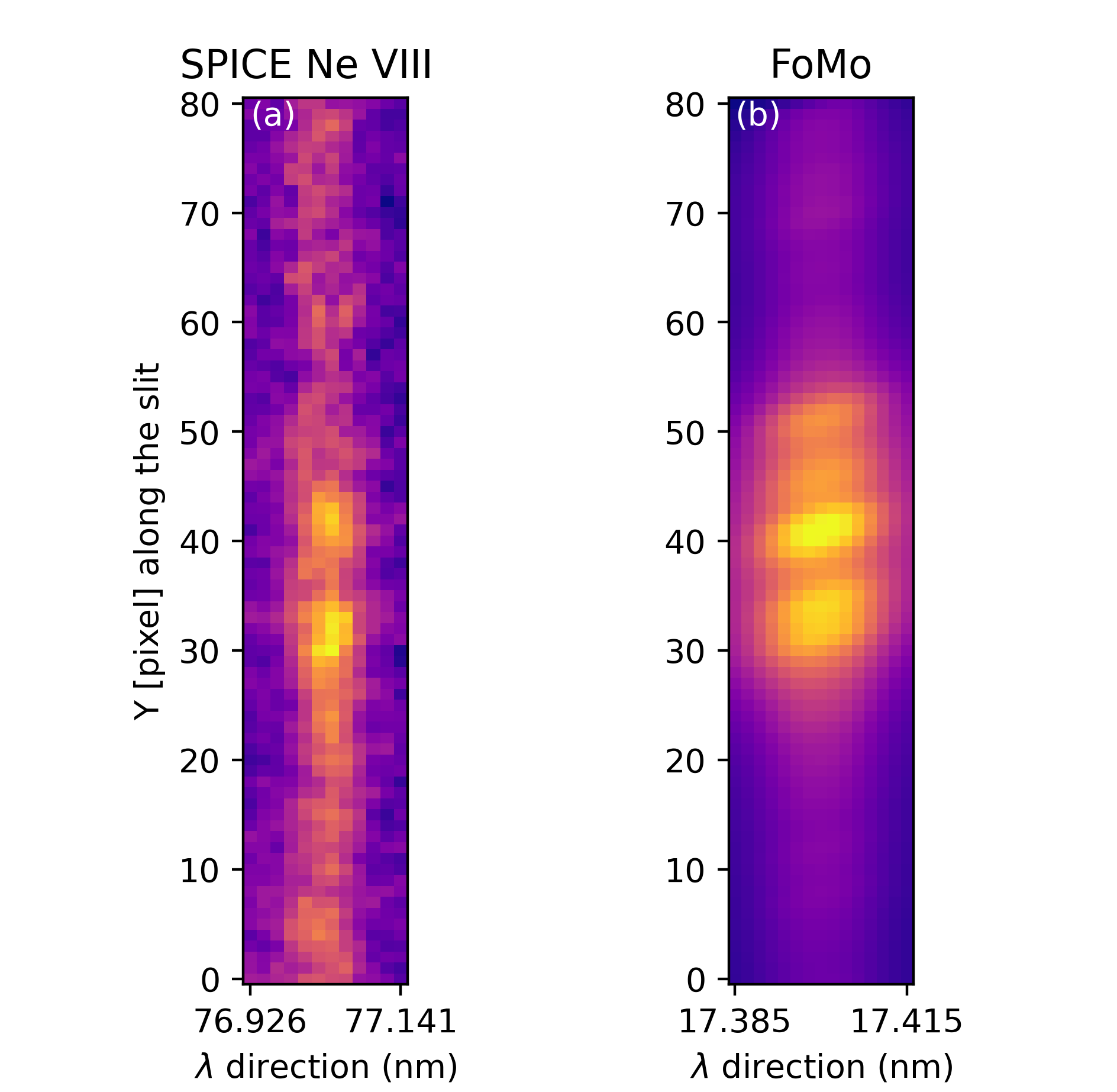}}
  \caption{Spectra from a particular slit position in SPICE and forward modelled data. The slit is passing through the spine region. }
  \label{fig:spectra}
\end{figure}

Fig. \ref{fig:spectra} displays the spectra for a specific slit position in SPICE, as illustrated in panel (a), and the modeled data depicted in panel (b). The slit is chosen to pass through the spine region. The intensity maps were aligned to ensure that the slit position traverses the same intensity feature. However, differences in spectra still persist because the channels depict different plasma temperatures.This discrepancy is the reason for the varying number of blobs observed along the slit. Nevertheless, the presence of rotation and elongation in the PSF is evident, particularly in \ref{fig:spectra} (b).

\section{Proxy for the instrumentally generated velocity }

An alternative way to check if the signal in SPICE Doppler maps is generated by the PSF tilt, is to estimate intensity gradients as it is done by \cite{Janvier2023}. False velocity signals exhibit correlations with image intensity gradients, particularly those aligned with the slit direction, with a smaller contribution from intensity gradients perpendicular to the slit. 
We follow the procedure described by \cite{Janvier2023} in Appendix B and construct intensity gradients for the intensity map obtained by SPICE and shown in Fig. \ref{fig:spice_intensity} (a).  
Next, the intensity gradient in each pixel is plotted against the estimated Doppler velocity in that pixel and a linear fit is performed. However, it is important to note that the linear fit is not necessarily the best choice due to the high level of spreading of the points. Fig. \ref{fig:fov} (a) shows negative correlation between Doppler velocity obtained with SPICE and intensity gradient along the slit similar to \cite{Janvier2023}. Contrary to the results of the results of \cite{Janvier2023}, Fig. \ref{fig:fov} (b) shows relatively strong positive correlation between velocities and intensity gradients across the slit. The $\pm1\sigma$ region was not included in order to perform the linear fit, as it is assumed to have a physical origin. 

\begin{figure}
 \centering
\begin{tabular}{c}
\includegraphics[width=.7\linewidth]{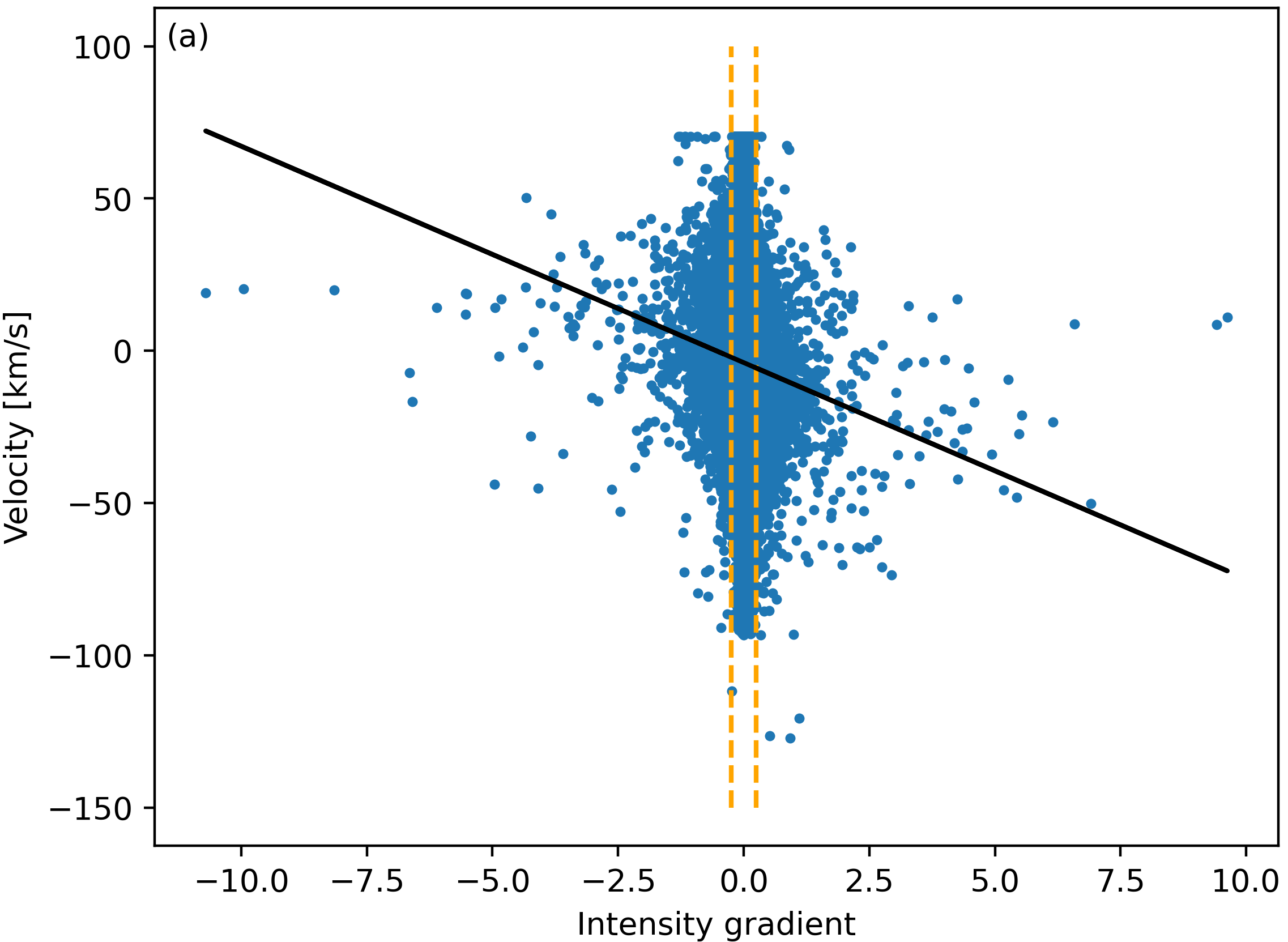}\\
\includegraphics[width=.7\linewidth]{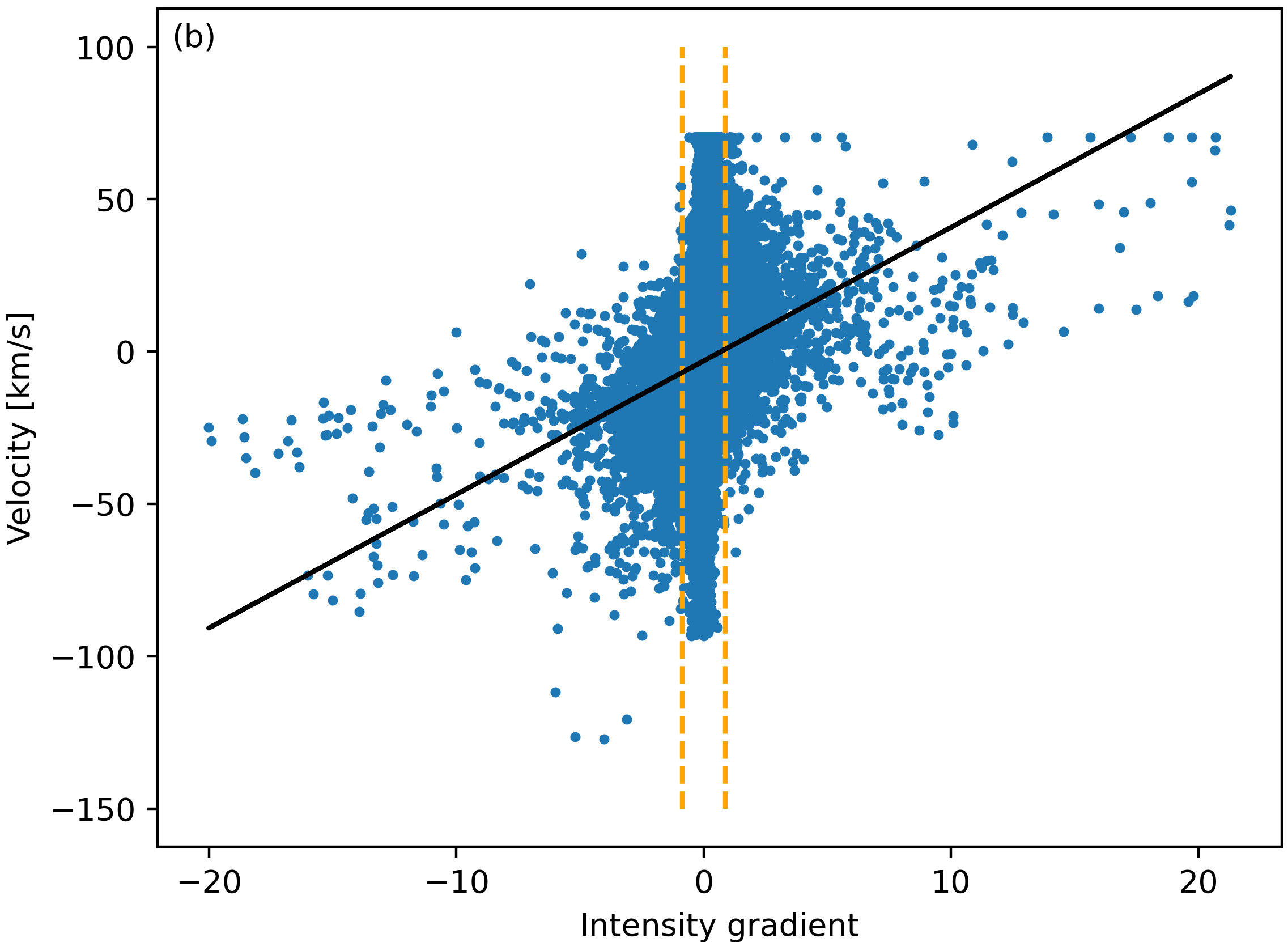}\\
\end{tabular}
\caption{Relation between the intensity gradients along (panel (a)) and across (panel (b)) the SPICE slit with the Doppler velocities. The orange dashed lines show the$\pm 1 \sigma$ range that was excluded from the linear fit (black line). This sample was taken from the C III raster at 14:02:59. }
\label{fig:correlation}
\end{figure}

\begin{figure}
  \resizebox{\hsize}{!}{\includegraphics{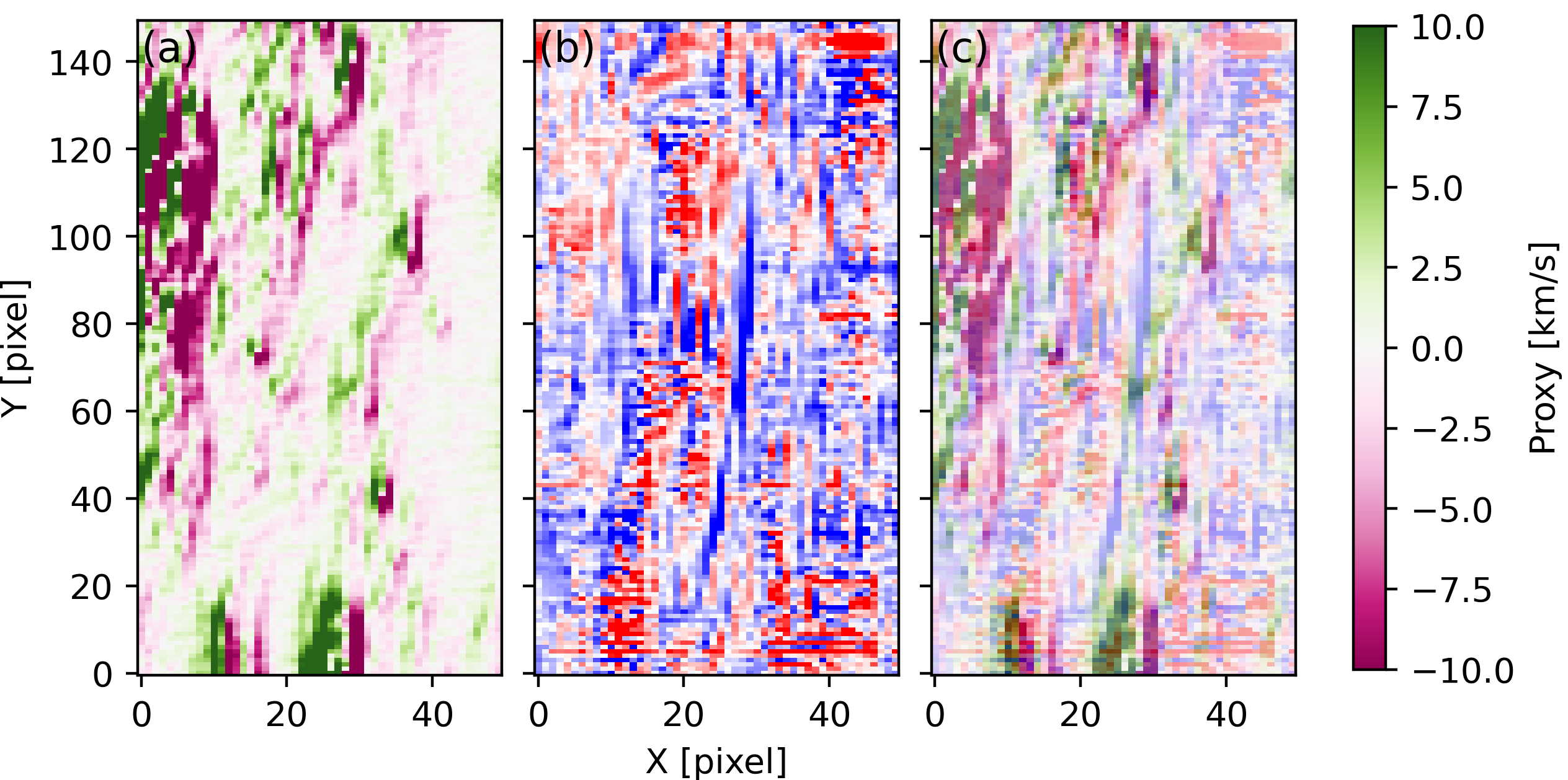}}
  \caption{Simulated Doppler velocity maps representing a proxy for the effect of SPICE PSF tilt on velocities (panel (a)). Panel (b) shows the Doppler velocities obtained with SPICE for comparison. Panel (c) shows simulated velocity maps on top of Doppler maps obtained with SPICE. }
  \label{fig:proxy}
\end{figure}

Finally, the slopes of the line fits were multiplied by the intensity gradients, and the resulting maps were summed together to create an overall proxy map shown in Fig. \ref{fig:proxy} (a) (which also has units of velocities). Here the white regions correspond to the areas with the minimal risk of getting false signal. 

Panel (b) shows the Doppler map obtained with SPICE taken with the C III raster at 14:02:59. The proxy map serves the purpose of identifying regions at greater risk of producing artificial signal. Fig. \ref{fig:proxy} (c) shows the proxy map imposed on top of the Doppler map allowing to match the location of spine with the proxy map. 
From panel (c) we can deduce that the spine region is sufficiently secure because it does not coincide with the high-risk areas for acquiring false signals. 

Consequently, these findings serve as additional evidence that supports the results obtained through forward modeling analysis.

\end{document}